\documentclass[manuscript,screen,nonacm]{acmart}

\setcopyright{none}
\usepackage{algorithmic}
\usepackage{textcomp}
\usepackage{xcolor}
\usepackage{comment}
\usepackage{graphicx,pstricks}
\usepackage{graphics}
\graphicspath{{img/}}
\usepackage{booktabs}
\usepackage{tabularx}
\usepackage{xspace}
\usepackage{caption}
\usepackage{array}
\usepackage{cancel}
\usepackage{makecell}
\usepackage{multirow}
\usepackage{enumitem}
\usepackage{colortbl}
\usepackage{subfig}
\usepackage[scaled=0.85]{beramono}
\newcommand{\cmark}{\checkmark}
\newcommand{\xmark}{$\times$}

\usepackage[dvipsnames]{xcolor} % For custom colors
\usepackage[most]{tcolorbox} % For creating colored boxes

\definecolor{lightgreen}{rgb}{0.894, 0.961, 0.949}
\definecolor{darkgreen}{rgb}{0.0, 0.576, 0.533}

\tcbset {
  base/.style={
    enhanced,
    breakable,
    arc=0mm, % Square corners
    boxrule=0mm,
    colback=lightgreen!20!, % Background color
    left=3.5mm,
    leftrule=2mm, % Thickness of the left stripe
    right=3.5mm,
  }
}

\newtcolorbox{mainbox}[1]{
  colframe=darkgreen, 
  base={#1}
}
\newcommand{\ie}{\emph{i.e.,}\xspace}
\newcommand{\eg}{\emph{e.g.,}\xspace}
\newcommand{\etal}{\emph{et al.}\xspace}

\AtBeginDocument{%
  }

\begin{document}

\title[Benchmarking the Quality Gap Between Human and AI Code]{What is the Difference Between Me and You? Benchmarking the Quality Gap Between Human-Written and AI-Generated Code}

\author{Cristina Improta}
\affiliation{%
  \institution{University of Naples Federico II}
  \city{Naples}
  \country{Italy}}
\email{cristina.improta@unina.it}

\author{Pietro Liguori}
\affiliation{%
  \institution{University of Naples Federico II}
  \city{Naples}
  \country{Italy}}
\email{pietro.liguori@unina.it}

\author{Domenico Cotroneo}
\affiliation{%
  \institution{University of North Carolina at Charlotte}
  \city{Charlotte, NC}
  \country{USA}}
\email{d.cotroneo@charlotte.edu}

\begin{abstract}
AI coding assistants are becoming co-authors of production software, yet their evaluation centers on functional correctness, leaving open whether their code differs from human code in the quality dimensions dominating lifecycle cost. We compare human-written and AI-generated code at scale: 787,562 function pairs across Python, Java, and C, each human function mined from open-source repositories paired with implementations generated from its docstring by three AI assistants (OpenAI GPT models, DeepSeek-Coder, Qwen2.5-Coder). We characterize structural complexity and statistical naturalness, and map static-analysis findings onto Orthogonal Defect Classification for defects and the Common Weakness Enumeration for vulnerabilities, making authors and languages directly comparable. AI-generated code is structurally compressed and stylistically templated: roughly half the size and branching of human code, clustering apart at the style level. Defect profiles differ in kind: human code concentrates issues of mature codebases, AI code repetitive boilerplate; security is language-dependent, with LLMs producing more, and more severe, findings in Python and Java but fewer high-severity memory-safety findings than humans in C. Once size is controlled for, complexity metrics carry little signal, while naturalness separates authors. Finally, we release \emph{CQBench}, a benchmark of 27,346 issue-prone tasks with baselines and an evaluation pipeline for quality assurance and security testing.
\end{abstract}




\maketitle
\begingroup
\renewcommand\thefootnote{}%
\footnotetext{This article is a revised and extended version of a paper presented at the 36th IEEE International Symposium on Software Reliability Engineering (ISSRE 2025)~\cite{cotroneo2025human}.}%
\addtocounter{footnote}{-1}%
\endgroup

\section{Introduction}
\label{sec:introduction}
AI coding assistants have moved from research prototypes to a routine part of software development in the span of a few years. Tools built on large language models (LLMs) now autocomplete, refactor, and synthesize entire functions and classes inside mainstream editors and continuous integration pipelines. Their appeal is immediate: they can reduce the effort required to produce an initial implementation, accelerate prototyping, and make programming assistance available directly inside development environments. A growing share of the code in open-source repositories and production systems is already authored, or co-authored, by these models rather than by \textit{people}~\cite{liu2026debt, octoverse2025, nadella2025llamacon}.

As software development shifts from a human-driven activity to a model-driven one, the developer's role changes with it, from writing to reviewing and integrating code produced by a system whose failure modes are not yet well understood. Increasingly, developer effort moves upstream of the code itself, from authoring statements to specifying intent through prompts and then revising the generated artifact. 

Driven by the traction these tools have gained, research on these tools has concentrated on two questions: whether the generated code is functionally correct, and whether it makes developers faster. Much of the literature has therefore evaluated AI-generated software through the lens of productivity~\cite{ziegler2022productivity} or functional correctness, asking whether a generated program compiles, passes tests, or solves a benchmark task~\cite{chen2021evaluating,austin2021program,hendrycks2021measuring,yu2024codereval}. These criteria are important, but they capture only a specific moment in the software lifecycle: the moment code is written.

The cost of software, however, is not dominated by its initial writing: the consequences of AI-assisted software development extend well beyond code production. Software must be read, reviewed, tested, debugged, secured, extended, and maintained over time, and the bulk of total lifecycle cost falls in this maintenance and evolution phase rather than in first authorship~\cite{lientz1980software, iso25010}.
Non-functional quality attributes such as maintainability, comprehensibility, reliability, structural complexity, and security are therefore central to the long-term cost of software evolution and to the accumulation of technical debt~\cite{avgeriou2016managing}. 
This concern is not hypothetical for AI-assisted development: a recent large-scale study of AI-authored commits in real repositories finds that AI-generated code routinely introduces code smells, correctness, and security issues, and that a substantial fraction of these issues are never fixed, persisting and accumulating as technical debt over the life of the repository~\cite{liu2026debt}.
A function that is generated quickly and satisfies the visible tests may still be difficult to modify, rely on brittle assumptions, omit relevant checks, or introduce vulnerabilities that become costly only after integration, producing defect patterns, security exposure, and maintenance debt that emerge only after generated code becomes part of real systems.

This sets up a concern that correctness-centric evaluation cannot address. If LLMs produce code that is qualitatively different from code written by human developers, not merely as correct or less correct, but differently structured, differently styled, or differently vulnerable, then the immediate productivity benefits of AI-assisted development may come bundled with hidden future costs. 

These consequences manifest along three main dimensions. 
First, LLM-generated code is increasingly committed to open-source projects and production systems with limited inspection, where it is then read, reused, and extended by human developers and, in turn, ingested as training data for the next generation of models; a systematic quality bias in this code therefore does not stay local, as insecure patterns propagate into deployed software and widen the attack surface. 
Second, the effects are distributional and build up over time: structurally impoverished code that omits error handling and edge cases fails in ways unit tests do not catch, and repetitive, templated output erodes the stylistic and structural diversity of the codebases it enters, accumulating maintenance burden that surfaces only during later evolution. Because these effects are slow and diffuse, they are largely invisible to the authoring-time, per-task evaluation that dominates the literature, yet they are precisely the costs that determine whether AI-assisted development is a net gain over a software system's lifetime. 
Third, the quality assurance practices, review heuristics, and tooling currently in use were calibrated on human-authored code; if AI-authored code departs from that baseline in systematic ways, those defenses are aimed at the wrong targets, and the developers now acting as reviewers rather than authors are left supervising a class of error they were not trained to anticipate or even notice (\eg hallucinated APIs or libraries).

Establishing whether such a gap exists, and characterizing its shape, therefore requires a systematic, large-scale comparison of the quality of human-written and AI-generated code.
Recent work has begun to broaden the evaluation of AI-generated code beyond functional correctness. Multi-dimensional benchmark studies evaluate generated solutions in terms of reliability, maintainability, complexity, code smells, and
security-focused studies that analyze vulnerabilities via static analysis
or repository-mined snippets~\cite{liu2024refining,li2026multi,pearce2025asleep}. These studies provide important evidence about the risks and limitations of AI coding assistants, but their findings are difficult to compare directly: they rely on different benchmarks, languages, static-analysis tools, metrics, and study-specific notions of quality. The literature thus remains fragmented, offering many useful observations but few standardized points of comparison between human-written and AI-generated code. Section~\ref{sec:related} surveys this body of work in detail.

This work takes a complementary perspective. Rather than asking whether an LLM can solve a programming task, we ask how the quality of AI-generated code differs from human-written code when both are analyzed at scale and under comparable conditions. 
Meaningful comparison requires abstracting away from tool-specific output and onto established, tool-independent taxonomies. We therefore anchor the analysis in two standardized frameworks with a long track record in software engineering: \emph{Orthogonal Defect Classification} (ODC) for defects and the \emph{Common Weakness Enumeration} (CWE) for security vulnerabilities. Mapping the heterogeneous outputs of language-appropriate analyzers onto these shared taxonomies is what makes multiple authors and languages directly comparable in the same study. 
% \domy{Alla luce dei nostri related work non e' piu' cosi'
% To our knowledge, no prior study has carried out this comparison at large scale across multiple programming languages and multiple quality dimensions at once, under a single established classification.
% Table 1 mostra Mao et al. 2026 con H-vs-AI, Multi-dim, Real, Large, Multi-lang tutti spuntati. Dobbiamo fin d'ora puntare a  taxonomy-grounding (ODC+CWE come strato comune) più il benchmark, non la combinazione generica. Restringi il claim su quest ultimo putno.
% }

We consider nearly 800k human-written code functions sourced from more than 34k real-world open-source GitHub repositories and use the extracted code docstrings to produce an AI-generated implementation of each function using three state-of-the-art AI coding assistants (\ie the OpenAI GPT family~\cite{chatgpt}, DeepSeek-Coder~\cite{guo2024deepseek}, and Qwen2.5-Coder~\cite{hui2024qwen2}), across three programming languages chosen to span distinct paradigms, typing disciplines, and application domains: Python, a dynamically-typed scripting language dominant in AI/ML and data work; Java, a statically-typed object-oriented language prevalent in enterprise back-ends where maintainability is a first-order concern; and C, a low-level systems language whose manual memory management exposes a class of safety issues absent from managed languages. These three languages let us test whether the observed trends generalize beyond a single programming paradigm. 
For each author and language we characterize three complementary aspects of software quality: \textit{(i)} structural complexity and statistical naturalness, which capture how code is organized and how stylistically predictable it is; \textit{(ii)} defects, classified through Orthogonal Defect Classification (ODC); and \textit{(iii)} security vulnerabilities, classified through the Common Weakness Enumeration (CWE). We then ask whether the first set of properties helps explain the second.

\paragraph{\textbf{Findings.}}
Across all three languages, AI-generated code is structurally compressed and more templated than human code: every model produces functions with roughly half the lines, half the cyclomatic complexity, and a third to a half of the Halstead volume of human functions, and a software naturalness analysis confirms that machine-generated code clusters together and apart from human code at the style-level. The defect and security profiles differ in kind, not only in degree: human code concentrates the issues characteristic of mature, evolved codebases, while AI code concentrates simpler, repetitive patterns such as unused parameters, and skews toward distinct, often higher-severity vulnerability classes. These patterns are graded across models and, for security, reverse direction in C, where human-written code carries the larger share of high-severity findings. The ``simpler but riskier'' characterization of AI code thus holds broadly but is language- and model-dependent, which argues for language-specific quality assurance rather than a single cross-language risk model.
A demonstration evaluation of a frontier model released after the study (Claude Opus 4.8) confirms that the identified quality risks are not confined to the studied generators: on a 600-task subset of our benchmark, the model still produces defects on roughly two thirds of the tasks and security findings on roughly one third, concentrated in injection and concurrency weaknesses, and passes the strict clean gate only on a minority of tasks in every language.

This paper makes the following contributions:
\begin{itemize}
    \item \textbf{A large-scale, multi-language characterization of human-written and AI-generated code quality.}
    We compare human-written code with code generated by multiple LLMs across Python, Java, and C, spanning multiple software quality aspects (\ie maintainability and comprehensibility, reliability and security) over languages with different paradigms, type systems, and quality-risk profiles.

    \item \textbf{A corpus-level analysis of code structure, style, and naturalness.}
    We characterize human and AI authors using a set of established software metrics that capture complementary aspects of code size, control-flow complexity, lexical diversity, maintainability, and naming, and we complement them with naturalness analyses based on cross-entropy and perplexity of code. 

    \item \textbf{A taxonomy-grounded analysis that makes heterogeneous results comparable.}
    We combine language-appropriate static analyzers with established classification frameworks: Pylint, PMD, and Clang-Tidy findings are mapped to ODC defect types, while Semgrep security findings are analyzed through CWE identifiers and severity levels. This design enables comparison across tools, languages, and code authors without relying on raw analyzer-specific findings.

    \item \textbf{A cross-language analysis of quality signatures and their relationships.}
    We examine whether the same ``simpler but riskier'' patterns generalize across Python, Java, and C, and we study correlations between structural and complexity metrics and defect and vulnerability profiles.

    \item \textbf{A public benchmark and evaluation framework for future studies of generated-code quality.}
    We release \emph{CQBench}, a public benchmark and evaluation framework for AI-generated code quality, including a curated set of the most issue-prone (``hardest'') functions and a pipeline that reproduces our analyses and scores new code against them. This enables future comparisons of additional LLMs and quality-assurance techniques under the same evaluation pipeline. We demonstrate the framework end-to-end on a frontier model released after the benchmark's construction (Claude Opus 4.8), showing that the curated tasks remain challenging: the model's completions trigger the same classes of defects and vulnerabilities the tasks were selected for~\cite{replication}.
\end{itemize}

This article extends our ISSRE 2025 conference paper~\cite{cotroneo2025human}, which compared human-written and AI-generated code in Python and Java with respect to defects, vulnerabilities, and structural complexity. The journal version substantially expands both the empirical scope and the analytical and methodological depth of the study.
\textit{First}, it adds C as a third programming language, requiring the construction of a new dataset, the integration of a new language-specific defect analyzer, a new rule-to-ODC mapping, and a new set of language-specific results. C is not merely a third data point: as a low-level systems language with manual memory management, it differs from Python and Java in paradigm, type discipline, and dominant risk profile, and its inclusion yields conclusions that could not be drawn from the conference version.
\textit{Second}, it broadens the structural analysis by adding further complexity and maintainability metrics, allowing a more detailed characterization of code size, control flow, lexical structure, and maintainability.
\textit{Third}, it introduces an entirely new naturalness analysis based on $n$-gram language models to capture stylistic and statistical differences that are not visible from conventional complexity metrics alone.
\textit{Fourth}, it deepens the defect and security characterizations: the per-author analysis of the most frequent specific defects is extended to C and consolidated into systematic cross-language rankings, and the security analysis adds an entirely new severity-stratified view of CWE findings that separates high-severity, directly exploitable patterns from context-dependent warnings.
\textit{Fifth}, it adds a new correlation analysis that relates structural and statistical properties to defect and vulnerability outcomes, assessing whether complexity and naturalness help explain quality differences across authors and languages.
\textit{Finally}, it releases a public benchmark and evaluation framework, including the curated corpus, analysis scripts, taxonomy mappings, security configurations, and aggregated results, to support replication and future studies of AI code quality, and demonstrates that the benchmark remains challenging for frontier AI coding assistants.

The remainder of this article is organized as follows.
Section~\ref{sec:related} positions the paper and reviews related work;
Section~\ref{sec:research_study} describes the research study and analysis
pipelines;
Section~\ref{sec:dataset} details the dataset construction process;
Section~\ref{sec:setup} describes the experimental setup;
Section~\ref{sec:evaluation} presents the results;
Section~\ref{sec:benchmark} showcases the benchmark constructed in this work;
Section~\ref{sec:discussion} discusses our findings and practical implications;
Section~\ref{sec:threats} details threats to validity, and
Section~\ref{sec:conclusion} concludes.

\section{Related Work}
\label{sec:related}
Research on AI-generated code has primarily evaluated whether generated programs satisfy a given specification. Benchmarks such as HumanEval~\cite{chen2021evaluating}, MBPP~\cite{austin2021program}, APPS~\cite{hendrycks2021measuring}, and CoderEval~\cite{yu2024codereval} have made this evaluation practical by pairing programming tasks with reference implementations or test suites in different languages. Such benchmarks are useful for measuring problem-solving ability, but they only partially characterize the \emph{quality} of the software that AI coding assistants generate. Code that is functionally correct, \ie passes the available tests, may still be difficult to maintain, stylistically atypical, vulnerable when deployed, or prone to defects and smells.

Several studies therefore extend benchmark-based evaluation beyond functional correctness by considering additional software-quality dimensions. Liu \etal~\cite{liu2024refining} combine execution-based correctness, static analysis, and a taxonomy of quality issues, showing that test-passing ChatGPT solutions may still contain maintainability problems. Liu \etal~\cite{liu2023your} strengthen HumanEval test suites and show that weaker tests can overestimate correctness. Li \etal~\cite{li2026multi} scale this perspective to 65,325 suggestions produced by five AI coding assistants across five languages, evaluating correctness, reliability, maintainability, and complexity. Other work combines tests with static-analysis signals to study smells, maintainability, reliability, security, and repair loops~\cite{tosi2024studying,blyth2025static,paul2025investigating,sabra2025assessing}.

These studies establish that correctness is not enough, but they also illustrate how fragmented the current evidence is.
Many still evaluate AI-generated code in isolation, on benchmark tasks, on small datasets, or through tool-specific quality indicators. They therefore leave only partially addressed a complementary question: whether code generated by LLMs differs systematically from human-written code when both are analyzed under comparable conditions. This question matters because quality-assurance tools, defect taxonomies, and maintainability metrics have historically been calibrated on human-authored software. If AI-generated code differs structurally, stylistically, or in its defect and vulnerability profiles, then quality assessment for AI-assisted development requires evidence that goes beyond raw performance, pass rates, and isolated security warnings.

More closely related to our positioning are studies that address one or more of these dimensions explicitly. Zheng \etal~\cite{zheng2024beyond} and CWEval~\cite{peng2025cweval} propose reusable benchmark or evaluation frameworks for assessing code generation beyond plain functional correctness, with CWEval focusing specifically on functionality and security. Kharma \etal~\cite{kharma2026security} study security and quality across multiple languages and models, while Sabra \etal~\cite{sabra2025assessing} quantify bugs, code smells, and vulnerabilities in Java code generated by contemporary LLMs. Other studies move closer to the human-vs-AI setting: Patel \etal~\cite{patel2024comparative} compare AI-generated and human-written Java code using software metrics and defect findings, Lee \etal~\cite{lee2026human} compare insecure uses of security-sensitive functions in AI-generated and human-written C code, and Jamil \etal~\cite{jamil2025can} assess GPT-generated solutions against human-written HumanEval baselines using code-quality metrics. Finally, Tambon \etal~\cite{tambon2025bugs} characterize recurring bug patterns in LLM-generated code through an explicit taxonomy, while Fu \etal~\cite{fu2025security} and Schreiber and Tippe~\cite{schreiber2025security} analyze security weaknesses in AI-attributed code mined from public repositories and organize findings through CWE-based classifications.

Most closely related, Mao \etal~\cite{mao2026large} present a large-scale study of AI-generated code mined from real-world repositories. Their study and ours differ in non trivial ways. First, they identify AI-generated code in the wild through a detection pipeline combining heuristic filtering and LLM-based classification, so authorship is inferred and no human implementation of the same functionality is available; we instead generate an AI implementation from the docstring of each human function, yielding function-for-function paired samples in which any observed difference is attributable to authorship rather than to task selection or repository norms. 
Second, their analysis centers on commit-level process characteristics such as change size, post-commit evolution, and collaboration patterns, reporting raw per-KLOC issue densities; we instead characterize the code's intrinsic quality signature through complexity, naturalness, and defects and vulnerabilities mapped onto the established ODC and CWE taxonomies, comparing defect and weakness types across languages and authors.
Third, their corpus covers managed languages (Python, JavaScript, TypeScript), while our inclusion of C surfaces the memory-safety vulnerability classes that dominate systems code and that we find behave opposite to the managed-language trend.

\tableautorefname~\ref{tab:rw} summarizes the studies most closely related to this paper. The table positions these studies along the dimensions jointly addressed here: direct human-vs-AI comparison, multi-dimensional quality evaluation, real-world repository code, large-scale analysis, multi-language coverage, explicit defect or vulnerability taxonomies, and reusable benchmarks, datasets, or evaluation frameworks.

\begin{table}[t]
\centering
\caption{Positioning of the most closely related studies against the dimensions jointly addressed in this work. \cmark{}~= addressed, \xmark{}~= not addressed. ``H vs AI'' = direct human-vs-AI comparison; ``Multi-dim'' = at least two quality dimensions beyond correctness; ``Real-world'' = real-world repository code; ``Large-scale'' = at least 10k code samples; ``Multi-lang'' = at least three languages; ``Taxonomy'' = findings organized through an explicit taxonomy, either standardized (\eg ODC, CWE) or study-specific; ``Benchmark'' = contributes a reusable benchmark, dataset, or evaluation framework.}
\label{tab:rw}
\small
\begin{tabular}{lccccccc}
\toprule
Study & H vs AI & Multi-dim & Real-world & Large-scale & Multi-lang & Taxonomy & Benchmark \\
\midrule
Zheng \etal 2024~\cite{zheng2024beyond}                     & \xmark & \cmark & \xmark & \xmark & \xmark & \xmark & \cmark \\
Patel \etal 2024~\cite{patel2024comparative}                & \cmark & \xmark & \xmark & \xmark & \xmark & \cmark & \xmark \\
Tambon \etal 2025~\cite{tambon2025bugs}                     & \xmark & \xmark & \cmark & \xmark & \xmark & \cmark & \xmark \\
Schreiber \& Tippe 2025 ~\cite{schreiber2025security}       & \xmark & \xmark & \cmark & \xmark & \cmark & \cmark & \xmark \\
Sabra \etal 2025~\cite{sabra2025assessing}                  & \xmark & \cmark & \xmark & \xmark & \xmark & \xmark & \xmark \\
Fu \etal 2025~\cite{fu2025security}                         & \xmark & \xmark & \cmark & \xmark & \xmark & \cmark & \xmark \\
Peng \etal 2025~\cite{peng2025cweval}                       & \xmark & \xmark & \xmark & \xmark & \cmark & \cmark & \cmark \\
Jamil \etal 2025~\cite{jamil2025can}                        & \cmark & \cmark & \xmark & \xmark & \xmark & \xmark & \xmark \\
R. Li \etal 2026~\cite{li2026multi}                         & \xmark & \cmark & \xmark & \cmark & \cmark & \xmark & \xmark \\
Kharma \etal 2026~\cite{kharma2026security}                 & \xmark & \cmark & \xmark & \xmark & \cmark & \cmark & \xmark \\
Lee \etal 2026~\cite{lee2026human}                          & \cmark & \xmark & \xmark & \xmark & \xmark & \cmark & \xmark \\
Mao \etal 2026~\cite{mao2026large}                          & \cmark & \cmark & \cmark & \cmark & \cmark & \xmark & \xmark \\
\midrule
\textbf{This work}                                          & \cmark & \cmark & \cmark & \cmark & \cmark & \cmark & \cmark \\
\bottomrule
\end{tabular}
\end{table}

The remainder of this section is organized around the software-quality dimensions examined in this study. Section~\ref{sec:rw-rq1} reviews structural complexity, style, and naturalness as proxy for \emph{maintainability and comprehensibility} of code; Section~\ref{sec:rw-rq2} reviews work assessing the \emph{reliability} of AI-generated software, including software defects and defect taxonomies; and Section~\ref{sec:rw-rq3} reviews \emph{security} vulnerabilities. Within each, we contrast benchmark-based evidence with the smaller body of human-vs-AI evidence and identify what a large-scale, multi-language, taxonomy-grounded comparison adds. We then close by examining the few studies that relate these dimensions to one another, which is the question our correlation analysis addresses: prior work has linked code naturalness to defectiveness~\cite{ray2016naturalness} and has shown that functional correctness and static-analysis quality can be largely orthogonal~\cite{sabra2025assessing}, but no study relates structural and statistical properties to defect and vulnerability profiles per author and per language on a paired human-AI corpus.

\subsection{Structural Complexity, Style, and Naturalness}
\label{sec:rw-rq1}

Structural metrics provide one way to compare human-written and AI-generated code beyond functional correctness. Lines of code, cyclomatic complexity, Halstead metrics and related measures capture complementary aspects of size, control-flow structure, lexical density, and maintainability. Prior benchmark-based evaluations use such metrics to quantify maintainability and complexity of generated solutions~\cite{li2026multi, sabra2025assessing}, while human-vs-AI studies use them to compare the structure of generated code against human baselines~\cite{patel2024comparative, cotroneo2025human}. These studies show that structural differences are not uniform across domains. For example, AI-generated code is often reported as shorter or less complex in general-purpose programming tasks, whereas AI-generated SQL has been found to be more structurally complex than human SQL queries under some prompt and model settings~\cite{Pecuchova_Benko_2026}. This suggests that the direction of the human-vs-AI structural gap depends on language, task, and prompting context.

Naturalness offers a complementary view. The naturalness hypothesis states that source code, like natural language, is repetitive and statistically predictable, making it amenable to language modeling~\cite{hindle2016naturalness}. Subsequent work refined this view by emphasizing localness in software~\cite{tu2014localness} and by showing that naturalness measurements can depend on tokenization and $n$-gram configuration~\cite{jimenez2018tuna}. Naturalness has been linked to software defects \cite{ray2016naturalness}, and more recent work uses statistical or neural representations to distinguish AI-generated from human-written code~\cite{xu2024one, suh2024empirical}. Detection studies show that AI code can carry identifiable stylistic signatures, but they also report generalization limits across datasets, languages, and generation settings.

The relationship between style, naturalness, and quality remains under-specified. Authorship-detection studies ask whether AI code can be recognized, not whether the recognized stylistic features are associated with maintainability, defects, or vulnerabilities. Conversely, complexity studies often quantify structural properties without measuring statistical predictability. Our study combines both perspectives: we measure structural properties and lexical diversity, and we use $n$-gram language models to quantify intrinsic naturalness, the contribution of identifier naming, cross-author transferability, and stylistic affinity. This allows us to characterize not only whether AI-generated code differs from human-written code, but also whether those differences are structural, lexical, statistical, or author-specific.

\subsection{Defects and Defect Taxonomies}
\label{sec:rw-rq2}

A second line of work examines the defects of LLM-generated code. Some studies derive taxonomies of recurring failure patterns from generated solutions. Tambon \etal~\cite{tambon2025bugs} analyze buggy LLM-generated functions and identify categories such as misinterpretations, missing corner cases, hallucinated objects, and wrong attributes. Dou \etal~\cite{dou2026wrong} extend this direction across several benchmarks and a real-world benchmark, reporting that functional bugs dominate and that generated solutions may be shorter while still exhibiting higher complexity than canonical implementations. Jahan and Wang \etal~\cite{jahan2025does} focus on erroneous assumptions made by GPT-4o, while Jesse \etal~\cite{jesse2023large} show that completion models can reproduce simple single-statement bugs and that such bugs may be statistically as natural as their fixes. 

These works are valuable because they move beyond aggregate pass rates and characterize the kinds of failures generated by LLMs. However, many taxonomies are study-specific, benchmark-specific, or language-specific. This makes cross-language comparison difficult: a Python runtime error, a Java static-analysis warning, and a C linting check may correspond to similar underlying defect types but appear under different tool-specific labels.

Orthogonal Defect Classification (ODC) provides a more general abstraction for comparing software defects across tools and languages \cite{chillarege1992orthogonal}. ODC classifies defects according to the nature of the corrective action required, rather than according to tool-specific rule names. Prior attempts to use ODC in this setting are limited in scope. For example, Bogaerts \etal~\cite{bogaerts2024taxonomy} apply ODC to Python vulnerabilities, indicating that defect taxonomies can support more interpretable comparisons. However, current literature does not provide a large-scale, multi-language view of human and AI defect profiles.

Our work uses ODC as the common layer over heterogeneous static analyzers. Rather than comparing raw static analysis rules directly, we map specific tools findings to ODC defect types and compare the resulting distributions across authors and languages. This design preserves language-specific analysis while enabling a shared interpretation of defect profiles.

\subsection{Security Vulnerabilities}
\label{sec:rw-rq3}

Security has received, alongside functional correctness, substantial attention in studies of AI-generated code. Scenario-based evaluations prompt models with security-relevant tasks and assess whether the generated code contains weaknesses. Pearce \etal~\cite{pearce2025asleep} evaluate GitHub Copilot on CWE-oriented scenarios and report that a substantial fraction of generated programs contain vulnerabilities. CWEval~\cite{peng2025cweval} evaluates functionality and security jointly through security-critical tasks and outcome-driven test oracles, showing that solutions that satisfy functional tests may still fail security requirements. These benchmark-based studies are important because they make security requirements explicit, but they typically analyze generated code without a matched human-written baseline. 

Repository-mining studies provide a different perspective by examining AI-attributed code in public projects. Fu \etal~\cite{fu2025security} mine Copilot-, CodeWhisperer-, and Codeium-attributed snippets from GitHub projects, map static-analysis findings to CWE categories, and evaluate whether prompt-based repair can remove weaknesses. Schreiber and Tippe~\cite{schreiber2025security} mine AI-attributed files from GitHub and use CodeQL to map vulnerabilities to CWE and severity information. These studies increase realism by observing code in repositories, but they do not compare each AI artifact against a human implementation of the same functionality.

Direct human-vs-AI security comparisons remain narrower. Jamil \etal~\cite{jamil2025can} compare HumanEval human baselines against GPT-generated solutions using code-quality metrics, including security-oriented tools. Sabra \etal~\cite{sabra2025assessing} analyze Java benchmark solutions with SonarQube and find that passing functional tests does not eliminate static-analysis issues, including bugs, smells, and vulnerabilities.

The security literature therefore provides strong evidence that AI-generated code may contain vulnerabilities, but it remains fragmented across security benchmarks, repository mining, and small human-vs-AI comparisons. Our study compares human-written and AI-generated code across Python, Java, and C, maps security findings to CWE categories, and analyzes severity levels and profiles. The goal is not just counting security warnings, but to determine whether human and AI authors exhibit different vulnerability profiles under a shared taxonomy to inform security testing practices.

In summary, existing work has provided substantial evidence on the functional correctness, maintainability, defectiveness, and security of AI-generated code. However, the literature remains fragmented across benchmark-based evaluation, human-vs-AI detection, defect characterization, and security analysis. To the best of our knowledge, no prior study jointly compares human-written and AI-generated code at large scale across multiple programming languages by combining structural complexity, code naturalness, ODC-based defect characterization, and CWE-based vulnerability analysis. Our study addresses this gap by characterizing not only whether AI-generated code differs from human-written code, but also how these differences manifest across quality dimensions relevant to long-term maintainability, technical debt, and security.

\section{Research Study}
\label{sec:research_study}
\begin{figure}
    \centering
    \includegraphics[width=0.8\linewidth]{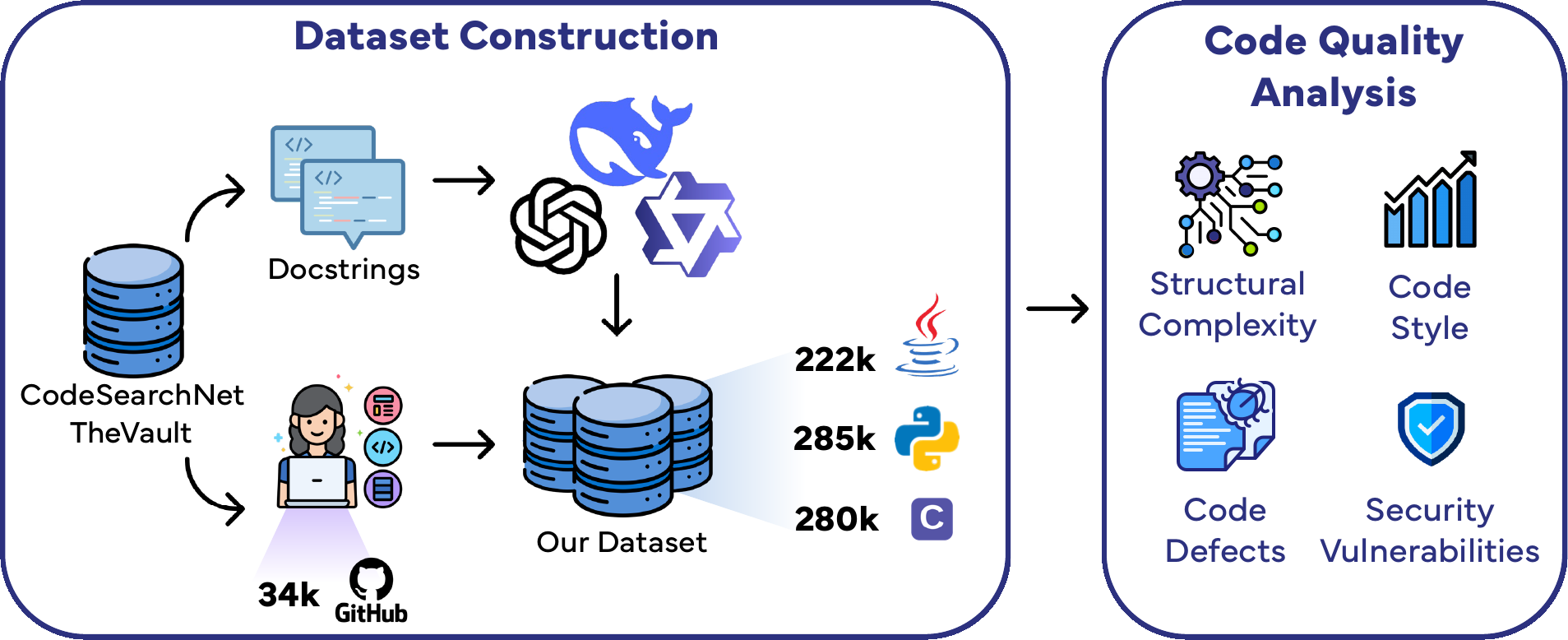}
    \caption{\textbf{Methodology overview.} Human functions from CodeSearchNet and TheVault (sourced from 34k GitHub repositories) are paired with AI implementations generated by three coding assistants from the original docstrings, producing 285k Python, 222k Java, and 280k C function pairs. Each pair is analyzed for structural complexity, style (\ie naturalness), defects, and security.}
    \label{fig:methodology}
\end{figure}

Comparing human-written and AI-generated code is a prerequisite for designing evaluation benchmarks, quality assurance practices, and tooling appropriate to AI-assisted development.
We designed this research study to assess differences between human-written and AI-generated code along multiple dimensions of software quality across programming languages that differ in paradigm, typing discipline, and domain of use. 

We focus on Python, Java, and C. Python is a dynamically-typed, high-level scripting language widely adopted in AI/ML and data-intensive workflows; Java is a statically-typed, object-oriented language predominant in enterprise and large-scale back-end systems, where maintainability and long-term support are primary concerns; C is a statically-typed, low-level systems language that underpins operating systems, embedded software, and performance-critical infrastructure, and whose manual memory management exposes a class of safety issues absent from managed languages. Together, these three languages span distinct ecosystems and quality requirements, allowing us to assess whether the observed trends generalize beyond a single programming paradigm. 

For each language, we consider code produced by human developers and by three state-of-the-art AI coding assistants (\ie ChatGPT, DeepSeek-Coder and Qwen2.5-Coder), selected for their prominence in recent code-generation benchmarks and their active use in software development workflows (see \S{}~\ref{sec:dataset} for details). 

To characterize the software produced by each \textit{code author}, we examine four complementary aspects of software quality: \emph{structural complexity}, which reflects size, control-flow intricacy, and lexical richness; \emph{statistical naturalness}, which captures the predictability and local regularity of code as a token sequence; \emph{defects}, which indicate violations of correctness, maintainability, and design principles; and \emph{security vulnerabilities}, which denote weaknesses exploitable by an adversary. Taken together, these dimensions provide a multi-faceted characterization of code quality that spans both internal attributes relevant to developers and external attributes relevant to the security and reliability of deployed software.

\subsection{Research Questions}

We designed this research study with the aim of answering the following research questions (RQs):
\vspace{0.1cm}

\noindent
$\rhd$ \textbf{RQ$_1$:} \textbf{Do code structural properties and style differ between human-written and AI-generated code across programming languages?}\\
To answer this research question, we characterize each corpus along two complementary dimensions. The first concerns the structural characteristics of the code, evaluated through a suite of function-level complexity and lexical-diversity metrics (\eg cyclomatic complexity, Halstead metrics), to examine whether AI-generated code shares similar surface-level and deeper structural traits with human-written code. The second concerns the statistical regularity of the code (its ``\textit{naturalness}''), quantified through cross-entropy and perplexity under $n$-gram language models. Together, these measurements assess both how code is organized and how stylistically predictable it is. All analyses are carried out across Python, Java, and C to determine whether the observed trends are consistent across languages.

\vspace{0.1cm}
\noindent
$\rhd$ \textbf{RQ$_2$:} \textbf{Do defect types and frequencies differ between human-written and AI-generated code across programming languages?}\\
To answer this research question, we systematically compare the distribution of defect types in human-written and AI-generated code across three programming languages. To this aim, we apply three state-of-the-art static analysis tools (\ie Pylint for Python, PMD for Java, Clang-Tidy for C) and map the resulting defects to standardized Orthogonal Defect Classification categories. This approach allows for a structured comparison of defect types and frequencies among \textit{code authors}, enabling us to quantify differences in code quality between human and AI-generated functions across Python, Java, and C datasets.

\vspace{0.1cm}
\noindent
$\rhd$ \textbf{RQ$_3$:} \textbf{Do security vulnerabilities differ between human-written and AI-generated code across programming languages, in terms of type and severity?}\\
The goal of this RQ is to investigate and compare security vulnerabilities present in human-written and AI-generated code across Python, Java and C code. Using Semgrep for static security analysis, we map the identified vulnerabilities to the Common Weakness Enumeration taxonomy and categorize them by severity level. This analysis enables us to assess differences in the types and criticality of security issues introduced by human developers versus AI code assistants, providing insight into potential real-world risks and \textit{author}-specific security behavior.

\vspace{0.1cm}
\noindent
$\rhd$ \textbf{RQ$_4$:} \textbf{Do the structural and statistical properties of code (RQ1) explain differences in defect and vulnerability profiles (RQ2, RQ3) across programming languages?}\\
The goal of this RQ is to assess whether the structural and statistical properties characterized in RQ$_1$ are associated with the defect and vulnerability outcomes measured in RQ$_2$ and RQ$_3$. To this aim, we compute per-language, per-author correlations between complexity and naturalness metrics on one side, and defect and vulnerability profiles on the other. This analysis enables us to determine whether and to what extent structural and stylistic properties are associated with differences in code quality, and whether these relations are stable across programming languages and code authors.

\subsection{Analysis Pipelines}

\subsubsection{Code complexity}
\label{sec:complexity-design}

To evaluate structural differences between human-written and AI-generated code, we measure a suite of code complexity metrics that are widely used in software engineering to assess maintainability, readability, and logical depth, summarized in Table~\ref{tab:metrics}. The metrics are organized into five complementary dimensions: \textit{(i)} size, \textit{(ii)} control-flow complexity, \textit{(iii)} lexical and data complexity, \textit{(iv)} maintainability, and \textit{(v)} one corpus-level metric capturing each author's vocabulary diversity across all generated code. The dimensions are deliberately complementary: size measures volume in isolation; control-flow metrics capture structural intricacy; the Halstead family~\cite{halstead1977elements}, which treats source code as a sequence of operators (keywords, punctuation, symbols) and operands (identifiers, literals), measures the data and lexical dimension that control-flow metrics miss; MI provides a composite per-function summary; and UT extends the analysis to the corpus level. This design follows recent evidence that no single complexity metric fully captures programmer comprehension difficulty, and that complementary combinations spanning size, control-flow, and Halstead families improve predictive accuracy~\cite{hao2025complementarity}. By analyzing these metrics, we aim to identify both quantitative differences between human and AI code, and stylistic or structural biases in how LLMs approach code generation. All metrics are computed per function except UT, computed once per author corpus. Operational details, including tools and configurations, are reported in \S{}~\ref{sec:setup-complexity}.

\begin{table}[t]
\caption{Structural complexity and lexical metrics, grouped by dimension. All
metrics are computed per function except Unique Tokens, computed once per author
corpus.}
\label{tab:metrics}
\small
\begin{tabular}{@{}llp{11.5cm}@{}}
\toprule
\textbf{Dimension} & \textbf{Metric} & \textbf{Definition} \\
\midrule
Size & NLOC & \textit{Number of Lines of Code.} Executable lines, excluding blanks
and comments; baseline indicator of verbosity. \\
\midrule
\multirow{6}{*}{\shortstack[l]{Control-flow\\complexity}}
 & CCN & \textit{Cyclomatic Complexity Number.} Linearly independent paths through
 the function; breadth of branching and decision-making. \\
 & MND & \textit{Maximum Nesting Depth.} Deepest control-flow nesting level; how
 deeply decisions are stacked, where CCN counts how many are made. \\
 & PC  & \textit{Parameter Count.} Formal parameters in the signature; broader
 interface surface and higher coupling to callers. \\
\midrule
\multirow{10}{*}{\shortstack[l]{Lexical and data\\complexity\\(Halstead~\cite{halstead1977elements})}}
 & $\eta_1$ & \textit{Distinct Operators.} Unique operators in the function, bounded
 by the language's syntax; breadth of language-level features. \\
 & $N_1$ & \textit{Total Operators.} Operator occurrences; with $\eta_1$, density
 and repetition of operator-level constructs. \\
 & $\eta_2$ & \textit{Distinct Operands.} Unique operands in the function; proxy for
 the local data vocabulary. \\
 & $N_2$ & \textit{Total Operands.} Operand occurrences; with $\eta_2$, how often
 each data item is referenced. \\
 & $V$ & \textit{Halstead Volume.} Information content from vocabulary size and
 token-sequence length; higher values indicate denser implementations. \\
 & $D$ & \textit{Halstead Difficulty.} Hardness to write or understand, increasing
 with operator variety and operand reuse intensity. \\
 & $E$ & \textit{Halstead Effort.} Product $V \times D$; unified estimate of the
 mental effort required to produce the code. \\
\midrule
\multirow{4}{*}{Maintainability}
 & MI  & \textit{Maintainability Index.} Bounded composite of $V$, CCN, and NLOC
 aggregating the size, control-flow, and lexical dimensions; higher values indicate
 code that is easier to maintain. \\
 & FNL & \textit{Function Name Length.} Character count of the function name; proxy
 for naming descriptiveness, widely associated with readability and long-term
 maintainability. \\
\midrule
Lexical diversity & UT & \textit{Unique Tokens.} Distinct lexical units across an
author's entire corpus; higher values indicate richer use of constructs and
identifiers, lower values more repetitive, template-like code. \\
\bottomrule
\end{tabular}
\end{table}

\subsubsection{Code naturalness}
\label{sec:naturalness-design}

We root this analysis in the \textbf{naturalness of software hypothesis}, first proposed by Hindle \etal~\cite{hindle2016naturalness} for human-produced software, and extend it to machine-generated code. 
According to this hypothesis, source code, like human language, exhibits strong statistical regularities because it is produced under syntactical constraints and in recurring contexts (\eg specific domains and functionalities). Consequently, it is highly repetitive and predictable, and this regularity can be captured by statistical language models and leveraged for a variety of software engineering tasks, including identification of buggy code~\cite{ray2016naturalness}, AI-generated software detection~\cite{xu2024one}, and adversarial code poisoning detection~\cite{sun2025show}. By applying statistical language models to sequences of code tokens authored by both humans and AI systems, we aim to quantify whether the predictability patterns that characterize human code persist or shift when generation is performed by LLMs.

Probabilistic metrics such as cross-entropy and perplexity are not only central to studies of naturalness but also lie at the foundation of today's AI code generation systems. Modern LLMs are themselves trained as next-token predictors, optimizing probabilistic likelihood over massive corpora of text and source code. Evaluating code with these same statistical measures therefore provides a direct lens into the patterns these models learn and reproduce. Both metrics quantify how \emph{surprising} a sequence of code tokens is under a learned distribution: \emph{cross-entropy} reflects the model's confidence in predicting each successive token, computed as the average negative log-likelihood of tokens under the model and expressed in bits per token; \emph{perplexity}, defined as the exponentiation of cross-entropy, translates this confidence into the model's effective branching factor, that is, the average number of equally probable continuations at any point in the sequence. Lower values of either metric indicate that the code follows familiar and coherent structural patterns, suggesting greater regularity. 

We compute both metrics using $n$-gram language models, which approximate the probability of a token from a bounded window of preceding context. Although large neural language models are a popular solution to score perplexity, $n$-gram models offer several practical and methodological advantages for the corpus-level analysis we conduct here. First, they are computationally tractable at million-function scale, where evaluating a code LLM on every function would be prohibitive. Second, $n$-gram perplexity is interpretable and model-agnostic: it depends only on the training corpus and the order $n$, rather than on a specific neural model's parameters or pretraining data, which is important when the entities being evaluated (LLM-generated code) overlap with the training corpora of any neural LM that might be used to score them~\cite{boreiko2025interpretable}. Third, despite their simplicity, $n$-gram models capture much of the local statistical regularity of source code, which is locally constrained by syntax and idioms~\cite{hindle2016naturalness}; recent work has further shown that suitably engineered $n$-gram models remain competitive with, and even complementary to, modern neural language models for perplexity-based analyses~\cite{liu2024infini}, and that for naturalness-based detection of perturbed code, simple $n$-gram models outperform code LLMs such as CodeBERT and CodeLlama in both accuracy and efficiency~\cite{sun2025show}. 

Operational details, including the tokenizer, the choice of $n$-gram orders, training protocol, and the language modeling toolkit, are reported in \S{}~\ref{sec:setup-naturalness}.

To assess whether there are stylistic and distributional differences in the code produced by human developers and by AI authors, we perform four complementary analyses designed to quantify and explain differences in naturalness via cross-entropy and perplexity. Together, these analyses assess \textit{(i)} how predictable each author's code is in its own style, \textit{(ii)} how much predictability is carried by naming versus structure, \textit{(iii)} how well stylistic regularities transfer across authors, and \textit{(iv)} how these relations organize into clusters of authorship.

\textbf{Is AI-generated software as ``natural'' as human-written software?}
Previous work found that code is repetitive and predictable, although to a varying degree across programming languages~\cite{hindle2016naturalness,rahman2019natural}. This first analysis aims to measure each author's intrinsic naturalness, \ie how repetitive and locally regular the code is. We establish a \textit{self-cross-entropy} curve for each author by training an $n$-gram language model, with $n$ ranging from 2 to 10, on a portion of that author's code and testing it on a held-out portion. Varying the order of $n$-grams, that is, the amount of local context the model can use, lets us assess the scale of predictability and whether regularities are driven mainly by short-range syntactic templates or by longer-range lexical and naming patterns.

\textbf{How much of code ``naturalness'' comes from lexical naming versus structural syntax?}
This second analysis isolates the contribution of naming practices, as opposed to code structure, by repeating the self-cross-entropy analysis on a normalized version of each snippet. The normalized representation preserves all structural tokens (keywords, delimiters, operators, parentheses, braces, and control-flow constructs) while systematically replacing identifiers and literals with role-aware placeholders for functions, classes, parameters, variables, attributes, imported modules, annotations, types, and literal values. This transformation collapses the otherwise large and sparse identifier vocabulary into a compact set of category tokens, reducing stylistic and semantic variation in naming while preserving the program's syntactic and structural form. As a result, self-cross-entropy values generally decrease under normalization. The magnitude of the decrease, \ie the gap between original and normalized values, indicates the extent to which naturalness depends on naming choices: larger gaps highlight strong reliance on identifier regularities, while smaller gaps suggest that predictability is dominated by structural patterns.

\textbf{To what extent do author-specific regularities transfer to others?}
The third analysis measures style specificity by testing cross-author transferability, \ie assessing whether an $n$-gram model trained on one author's code predicts another author's code well. Fixing $n$ at the order at which the self-cross-entropy curve plateaus, we train a language model on each author's code and evaluate it on the held-out test sets of every other author, computing the resulting \textit{cross-perplexity}. If code regularities are author-specific, a model should predict its own author's code more accurately than anyone else's, indicating locality of style. The same setup further allows us to compare how AI authors predict one another versus how they predict human code, and conversely.

\textbf{Can code authorship be distinguished by stylistic affinity?}
The fourth analysis synthesizes the pairwise modeling results into a global view by clustering authors using a \textit{cross-entropy-based distance} that captures how much harder it is to model someone else's code than one's own. Let $H_{X \to Y}$ denote the mean cross-entropy, in bits per token, of author $X$'s language model scoring author $Y$'s held-out code. The distance between two authors is the symmetrized excess cross-entropy
\begin{equation}
d(A,B) = \frac{1}{2}\left[\left(H_{A \to B} - H_{A \to A}\right) +
\left(H_{B \to A} - H_{B \to B}\right)\right],
\label{eq:excess_ce_distance}
\end{equation}
which subtracts each author's self-baseline so that the resulting values reflect only the additional difficulty of cross-author modeling, controlling for author-specific entropy levels (\eg code length, variability, or naming style); the distance is symmetric by construction and zero on the diagonal. Applying average-linkage agglomerative hierarchical clustering to the resulting distance matrix produces a hierarchy that organizes authors by stylistic affinity solely as a function of their excess cross-entropy distances, enabling a direct test of whether human-authored code separates from LLM-authored code, and whether subclusters emerge among specific LLMs. Dendrogram leaves are reordered for display via optimal leaf ordering, which affects neither the tree topology nor the merge heights.

\subsubsection{Code defects}

We systematically investigate software defects introduced by human developers and AI models when writing code. A \emph{defect} refers to any violation of correctness, logic, performance, or design. These include issues such as improper variable initialization, control-flow anomalies, faulty error handling, or structural design problems. They may lead to program misbehavior but do not necessarily expose the system to external threats.

To detect potential issues, we employ three widely-used static analysis tools, one per programming language: \textbf{Pylint}~\cite{pylint} for Python, \textbf{PMD}~\cite{pmd} for Java, and \textbf{Clang-Tidy}~\cite{clang} for C. These tools were selected because of their maturity, extensive adoption in industrial and academic software analysis workflows, and their ability to produce fine-grained, rule-specific outputs that can be aggregated and compared at scale. Each tool parses source code into an Abstract Syntax Tree (and, in the case of Clang-Tidy, also performs control-flow and dataflow analysis) and applies a set of rule-based patterns to detect coding flaws, such as syntax errors, type inconsistencies, uninitialized variables, error-prone constructs, performance bottlenecks, and design problems.

Pylint defines $\sim$400 detection rules covering syntax errors, type inconsistencies, uninitialized variables, and poor programming practices; PMD provides $\sim$300 rules organized into best-practice, code-style, design, documentation, error-prone, multithreading, performance, and security rule sets; Clang-Tidy provides hundreds of checks organized into modules (\eg \texttt{bugprone-*}, \texttt{cert-*}, and \texttt{performance-*}). Each rule (or \textit{check}, in Clang-Tidy terminology) carries a unique symbolic name, and findings include the rule identifier, source location, and a human-readable message.

In our setup, we exclude all rules concerning code style and documentation, such as formatting, naming conventions, and documentation warnings, as these do not directly affect code quality dimensions (\eg functional correctness, reliability). Additionally, we omit the security ruleset due to its limited coverage, opting instead to incorporate a more comprehensive, security-focused solution later on. 

The retained rule sets contain 350 rules for Python, 226 rules for Java, and 258 rules for C, as shown in \tableautorefname{~\ref{tab:odc-mapping}}, all of which target functional, structural, or design defects at the function level.

\paragraph{Mapping rules to a common taxonomy.}
To systematically categorize and compare software defects across different code authors and across different programming languages, we rely on the \textit{Orthogonal Defect Classification} (ODC) framework~\cite{chillarege1992orthogonal}. ODC is a widely-recognized method for classifying software defects based on their nature and origin, independent of the specific software development process or application domain. It is based on a set of attributes, which are non-overlapping dimensions that measure different aspects of defects, including their type, impact, trigger (\ie how the defect was found), and source. This orthogonality enables standardized analysis and consistent defect-pattern detection across heterogeneous tools and languages.

\begin{table*}[t]
\centering
\caption{Mapping between Orthogonal Defect Classification (ODC) defect types and rules in Pylint (Python), PMD (Java), and Clang-Tidy (C). Normal text indicates Python rules, \textbf{bold} text indicates Java rules, \textit{italic} text indicates C checks.}
\label{tab:odc-mapping}
\small
\begin{tabular}{@{}lp{4.2cm}rrrp{4.4cm}@{}}
\toprule
\textbf{ODC Defect Type} & \textbf{Defect Characteristics} & \textbf{Python} & \textbf{Java} & \textbf{C} & \textbf{Example Rules} \\
\midrule
Assignment              & Errors in assignment, initialization, or variable binding                                  & 46  & 13 & 38 & \texttt{used-before-assignment}, \textbf{\texttt{UnusedAssignment}}, {\footnotesize \textit{\texttt{bugprone-assignment-in-if-condition}}} \\
\addlinespace
Algorithm               & Logical flaws in computation or data manipulation                                          & 118 & 88 & 48 & \texttt{too-many-nested-blocks}, \textbf{\texttt{CyclomaticComplexity}}, \textit{\texttt{bugprone-infinite-loop}} \\
\addlinespace
Interface               & Issues with interaction between modules, functions, or APIs                                & 113 & 12 & 61 & \texttt{no-value-for-parameter}, \textbf{\texttt{UseProperClassLoader}}, \textit{\texttt{bugprone-posix-return}} \\
\addlinespace
Checking                & Faulty validation or error-handling mechanisms                                             & 23  & 26 & 33 & \texttt{missing-timeout}, \textbf{\texttt{EmptyCatchBlock}}, \textit{\texttt{cert-err33-c}} \\
\addlinespace
Timing/Serialization    & Concurrency, event ordering, or multithreading defects                                     & 1   & 11 & 19 & \texttt{useless-with-lock}, \textbf{\texttt{AvoidSynchronizedStatement}}, \textit{\texttt{concurrency-mt-unsafe}} \\
\addlinespace
Function/Class/Object   & Structural or design errors in function, class, or object organization                     & 49  & 76 & 59 & \texttt{redefined-outer-name}, \textbf{\texttt{MissingOverride}}, \textit{\texttt{misc-unused-parameters}} \\
\midrule
\textbf{Total}          &                                                                                           & \textbf{350} & \textbf{226} & \textbf{258} &  \\
\bottomrule
\end{tabular}
\end{table*}

In the paradigm of AI-assisted software development, where developers query LLMs with a short description of the required implementation, the generated code is usually composed of snippets, single functions, or short classes that are then integrated into a larger codebase. For this reason, not all ODC attributes are applicable to this fast-prototyping scenario. We therefore apply the \textit{Defect Type} attribute from ODC to categorize software issues identified in the code. Defect Type characterizes a fault by the corrective change required to remove it (\eg adding a missing check, fixing an assignment, restructuring an algorithm), independently of specific code details or language constructs and covers eight distinct categories. 

Following our previous work~\cite{cotroneo2025human}, we focus on the six categories that are meaningful at the function level: \emph{Assignment}, \emph{Algorithm}, \emph{Interface}, \emph{Checking}, \emph{Timing/Serialization}, and \emph{Function/Class/Object}. We deliberately exclude two categories from the analysis: \emph{Documentation}, because docstrings are stripped from code samples and repurposed as generation prompts during dataset construction, see \S{}~\ref{sec:dataset}, and \emph{Build/Package/Merge}, because these defects pertain to broader integration concerns such as library dependencies, version control, and packaging errors that are not applicable to isolated code snippets. 
We also exclude the ODC \textit{Trigger} attribute, which captures how a defect is discovered (\eg during inspection, unit testing, integration testing); in our setting, defect detection is performed entirely through static analysis applied uniformly to all samples, and assigning trigger labels would require speculative assumptions about a development workflow that does not exist for snippet-level LLM output.

To provide a shared taxonomy of code quality issues, we manually mapped each rule from Pylint, PMD, and Clang-Tidy to the most appropriate ODC \textit{Defect Type}. This mapping is necessary because static analysis rules are tool- and language-specific and cannot be directly compared across analyzers; ODC provides the abstraction layer that enables meaningful cross-language comparison. The mapping followed a structured three-stage protocol to mitigate subjectivity and reinforce reproducibility. First, an initial mapping was performed by one rater based on the diagnostic semantics of each rule and the kind of corrective action it implies. Second, the mapping was independently reviewed by two additional authors, including a domain expert. Third, all disagreements were resolved through discussion until consensus was reached. This procedure served as a form of inter-rater agreement and ensured conceptual alignment with the ODC taxonomy across the three tools.

The six retained categories cover distinct dimensions of software defects. \emph{Assignment} groups defects involving incorrect or missing variable initialization, value binding, or assignment, with corrective actions targeting the way values are set or initialized. \emph{Algorithm} captures logical or control-flow flaws whose correction requires re-implementing or restructuring the computation, such as fixing an incorrect loop condition or simplifying excessive branching. \emph{Interface} groups defects arising from incorrect interaction between a function and its callers, callees, or external APIs, with corrective actions involving the function's signature, return-value handling, or call-site arguments. \emph{Checking} captures faulty validation and error-handling, where the correction adds, removes, or restructures explicit checks on inputs, return values, or runtime conditions. \emph{Timing/Serialization} covers concurrency, event-ordering, and shared-resource defects, fixed by introducing or correcting synchronization, ordering constraints, or thread-safe alternatives. \emph{Function/Class/Object} groups structural and design defects in how a function, class, or data type is organized, with corrective actions reshaping the unit's structure rather than its internal computation.

To make the mapping process concrete, we discuss three representative examples, one per language. The Pylint rule \texttt{used-before-assignment}, which detects the use of variables before initialization in Python, is mapped to the \emph{Assignment} category, because the corrective action targets variable binding and initialization. The PMD rule \texttt{CyclomaticComplexity}, which flags Java methods with excessive branching logic, is classified under \emph{Algorithm}, as removing the defect requires restructuring control flow rather than fixing a single binding or check. The Clang-Tidy check \texttt{bugprone-posix-return}, which detects calls to POSIX functions whose return value is compared against the wrong sentinel (\eg comparing the return of \texttt{pthread\_create} against \texttt{-1} when the function returns a positive error number on failure), is classified under \emph{Interface}, as the correction aligns the call site with the function's documented return-value contract. These examples illustrate how each rule is assigned based on the functional semantics of the issue it detects and the corrective action it implies, rather than on tool-specific naming, ensuring consistency in our defect categorization across languages and tools.

\tableautorefname~\ref{tab:odc-mapping} summarizes the mapping by reporting, for each ODC \textit{Defect Type}, its defect characteristics, the number of mapped rules from each tool, and a representative example from each language. The complete rule-to-ODC mapping for all three tools is included in the replication package for transparency and future extensions~\cite{replication}.

By aggregating tool-specific findings into ODC categories, we obtain a structured, language-independent view of the defect profile of each code author, enabling direct comparison of how human developers and AI code assistants distribute their defects, highlighting differences in error types rather than being limited to tool output frequencies. 
Operational details, including the tool versions, the exact list of excluded rules, snippet-level handling are in \S~\ref{sec:setup-defects}.

\subsubsection{Code security}
\label{sec:security-design}

A \emph{security vulnerability} is a specific class of defect that poses a risk of exploitation, potentially allowing unauthorized access, data leakage, or system compromise; common examples include command injection, hardcoded credentials, and use of unsafe APIs.
To investigate and compare security vulnerabilities present in human-written and AI-generated code, we perform static security analysis on every function in the dataset and classify the resulting findings using a standardized vulnerability taxonomy.

We employ Semgrep OSS~\cite{semgrep}, a modern, lightweight static analysis tool designed for finding security vulnerabilities, correctness bugs, and code quality issues across more than 30 programming languages. Unlike traditional static analyzers that often require full code compilation (\eg CodeQL~\cite{CodeQL}, Bandit~\cite{bandit}), Semgrep operates by syntactic pattern matching directly on source code, making it particularly well-suited for scanning large-scale, heterogeneous datasets without requiring build artifacts or full type inference. 
Semgrep is widely deployed in industry as part of CI/CD pipelines and pre-commit security checks, with documented adoption across major technology, financial, and infrastructure companies, where it scans production codebases for vulnerabilities at scale~\cite{semgrep}. %This combination of academic and industrial use makes it a practical and well-validated choice for large-scale, cross-language vulnerability analysis. 

Semgrep provides an extensive registry of curated rule sets, each targeting specific families of issues, such as injection attacks, insecure communication, improper authentication, sensitive data exposure, and other widely recognized weakness categories aligned with security standards such as the CWE Top 25~\cite{top25mitre} and the OWASP Top Ten~\cite{owasp}. We apply Semgrep's curated security-focused rule sets uniformly across all three languages.

To enable a structured cross-language and cross-author comparison of vulnerabilities, we leverage Semgrep's native mapping of each rule to its corresponding entry in the \textit{Common Weakness Enumeration} (CWE)~\cite{cwe}. 
CWE provides a tool- and language-independent taxonomy of software weaknesses, organizing them into hundreds of named categories (\eg CWE-78 \emph{OS Command Injection}, CWE-798 \emph{Use of Hardcoded Credentials}). Each Semgrep rule carries a CWE identifier in its metadata, assigned and maintained as part of Semgrep's curation process; we use these identifiers as-is, without any manual reclassification. 

The complete configuration, including the list of security rules, are detailed in \S{}~\ref{sec:results-vulns} and made available in our replication package~\cite{replication}.

\subsection{Validation of Static Analysis Findings}
\label{sec:manual-validation}

Our quality measurements inherit the precision of the underlying analyzers. Exhaustive manual validation of findings is infeasible at the scale of this study, which spans 787,562 functions, four analyzers, and three languages, and prior work adopting a comparable design relies on the same trade-off~\cite{improta2025quality}. We mitigate this limitation in three ways. First, we employ mature, widely adopted tools (Pylint, PMD, Clang-Tidy, and Semgrep) whose rule sets are curated and maintained for industrial use, and whose findings have been manually validated with adequate precision in prior empirical studies~\cite{improta2025quality, fu2025security}. Second, we exclude entire rule families prone to subjective or context-dependent judgments (style, formatting, naming, and documentation rules), retaining rules that target functional, structural, or design defects. Third, we performed a randomized screening of findings across tools, languages, and code authors, inspecting the flagged code to confirm that reported issues corresponded to the patterns described by the triggering rules; this screening surfaced no systematic misfiring, although it does not constitute a statistical estimate of per-tool precision. More fundamentally, our conclusions do not rest on absolute finding counts: all comparisons are within-tool and within-language, so a rule's baseline false-positive rate applies uniformly to all four authors and cancels out of cross-author contrasts. Residual risk remains only where a rule's precision interacts with author-specific coding style; we flag the findings most exposed to this risk where they arise (\eg the \texttt{protected-access} convention in Python, \S~\ref{sec:results-defects}) and discuss the limitation in \S~\ref{sec:threats}.

\section{Dataset}
\label{sec:dataset}
\label{sec:dataset}

To investigate whether differences exist in the structural properties, naturalness, defect distribution, and security profiles of human-written and LLM-generated code, we curated a large-scale dataset spanning Python, Java, and C. We selected these three programming languages as the targets of our study due to their widespread adoption, differing programming paradigms, and complementary application domains~\cite{most-used-languages}. Python is a dynamically-typed, high-level scripting language heavily used in AI/ML and data-science development. Java is a statically-typed, object-oriented language widely deployed in enterprise-grade applications, where software quality and security requirements are stringent. C is a low-level, statically-typed systems-programming language used in operating systems, embedded software, and performance-critical infrastructure, where memory safety and low-level resource management dominate the defect landscape. By analyzing all three, we aim to ensure that our findings on code quality, defect profiles, and security vulnerabilities generalize across distinct language paradigms, type systems, and ecosystems.

The dataset is organized as a collection of $\langle$\emph{docstring}, \emph{human-code}, \emph{LLM-code}$\rangle$ tuples, where each LLM column corresponds to a different code-assistant model. For Python and Java, we extend the \textit{HMCorp} dataset of $\langle$\emph{human-code}, \emph{ChatGPT-code}$\rangle$ pairs released by Xu \etal ~\cite{xu2025distinguishing}, and add two additional state-of-the-art LLMs (DeepSeek-Coder and Qwen2.5-Coder) to enrich the AI-generated portion. For C, we build a new corpus starting from \textit{TheVault} dataset~\cite{nguyen2023vault}, on which we generate code with three different LLMs (OpenAI gpt-oss, DeepSeek-Coder, and Qwen2.5-Coder). \tableautorefname~\ref{tab:dataset_stats} reports the high-level composition of the dataset.

\begin{table}[t]
\caption{Dataset composition.}
\small
\label{tab:dataset_stats}
\begin{tabular}{
>{\raggedleft\arraybackslash}m{1.5cm}
>{\raggedleft\arraybackslash}m{2.5cm}
>{\raggedleft\arraybackslash}m{2.5cm}
>{\raggedleft\arraybackslash}m{2.5cm}
>{\raggedleft\arraybackslash}m{2.5cm}
}
\toprule
\textbf{Language} & \textbf{\# Samples} & \textbf{\# GitHub Repositories} & \textbf{Avg. Docstring Length} & \textbf{Avg. Code Length} \\
\toprule
\textit{Python} & 285,249 & 12,632 & 25.73 & 87.21 \\
\textit{Java}   & 221,795 & 4,296  & 24.86 & 79.30 \\
\textit{C}      & 280,518 & 17,139 & 39.42 & 81.51 \\
\bottomrule
\end{tabular}
\end{table}

\textbf{\textit{Python and Java.}}
We began by adopting and extending the \textit{HMCorp} dataset~\cite{xu2025distinguishing}, which contains 288,508 $\langle$\emph{human-code}, \emph{ChatGPT-code}$\rangle$ pairs in Python and 222,335 in Java. HMCorp was constructed by filtering the Python and Java subsets of the CodeSearchNet (CSN) dataset~\cite{husain2019codesearchnet} to extract functions authored by human developers. For each function, the corresponding docstring was used to prompt ChatGPT (``gpt-3.5-turbo'') to generate a matching AI implementation. During this filtering process, noisy or malformed samples, such as those containing HTML tags, URLs, or empty function bodies, were systematically removed to ensure dataset quality. The original CodeSearchNet dataset is a widely-used benchmark comprising $\langle$\emph{documentation}, \emph{code}$\rangle$ pairs in six different languages mined from public, non-forked GitHub repositories sorted by popularity (stars and forks), and is often used to train and evaluate LLMs on code-related tasks~\cite{zheng2023survey, wang2021codet5}.

The released HMCorp dataset, however, excluded the original docstrings and GitHub provenance. To restore this information, we performed a pre-processing step to match each HMCorp sample back to its source in CSN using a combination of function signature and body similarity. This step allowed us to recover both the documentation and repository metadata, resulting in the creation of complete $\langle$\emph{repository}, \emph{docstring}, \emph{human-code}, \emph{ChatGPT-code}$\rangle$ tuples for each instance. During this process, 3,259 Python and 539 Java samples could not be matched and were discarded, yielding a total of 285,249 Python and 221,795 Java instances drawn from 12,632 and 4,296 distinct GitHub repositories respectively.

\textbf{\textit{C.}}
For C, no equivalent of HMCorp exists. We therefore constructed a new corpus starting from \textit{TheVault}~\cite{nguyen2023vault}, a large-scale collection of $\langle$\emph{docstring}, \emph{code}$\rangle$ pairs %extracted from a subset of \textit{The Stack}~\cite{kocetkov2022stack} and 
curated to contain only high-quality function-level snippets in ten programming languages. TheVault provides 487,185 C samples partitioned into a small ($105,978$) and a medium ($381,207$) subset based on file-size criteria.

To match the docstring-quality and granularity assumptions made by HMCorp for Python and Java, we filtered TheVault's C subsets to retain only samples whose docstrings contained approximately 25 tokens or more, broadly comparable to the average docstring length of Python and Java functions. The resulting human-code corpus contains 280,518 unique C functions drawn from 17,139 distinct GitHub repositories, providing a representative mix of styles, domains, and development practices. Each retained sample preserves the original docstring (stripped of its comment delimiters), the function body, and the source-repository identifier, ensuring full traceability back to TheVault's release.

\subsection{Generation of LLM-Authored Code}
\label{subsec:dataset-generation}

To enrich the dataset with diverse AI-generated implementations, we employed three additional state-of-the-art code-generation LLMs alongside ChatGPT (already provided by HMCorp for Python and Java). Each model was prompted independently using the original docstring and corresponding function signature extracted from the human-authored code.

\noindent
$\blacksquare$ \textbf{DeepSeek-Coder-Instruct (33B)} (\emph{``DSC''})~\cite{guo2024deepseek} is part of a series of code language models, each trained from scratch on 2T tokens, with a composition of 87\% code and 13\% natural language. The base model is pre-trained on a project-level code corpus by employing a window size of 16K and an additional fill-in-the-blank task. We use a quantized, instruction-tuned variant with 33 billion parameters, further fine-tuned on 2B tokens of instruction data.\\
\noindent
$\blacksquare$ \textbf{Qwen2.5-Coder-Instruct (32B)} (\emph{``Qwen''})~\cite{hui2024qwen2} is the latest series of Qwen large language models specifically designed for code-related tasks. These models have been pre-trained on a dataset exceeding 5.5 trillion tokens, achieving state-of-the-art performance in code generation, completion, reasoning, and repair tasks. We employ a quantized model with 32 billion parameters, further trained for instruction-tuning through a multi-stage fine-tuning process.\\

\noindent
$\blacksquare$ \textbf{OpenAI GPT family.} For the OpenAI provider we use two models from the same lineage, one per language subset. For Python and Java we use ChatGPT (``\textbf{gpt-3.5-turbo}''), as provided by the HMCorp dataset~\cite{xu2025distinguishing}. For C, where HMCorp offers no implementations, we use \textbf{gpt-oss-20B}~\cite{gptoss}, the open-weight code-and-reasoning model released by OpenAI; we adopt the 20-billion-parameter variant, which targets cost-efficient inference while remaining within the same model family that produced ChatGPT and same size range as the other two adopted models. We report both under a single \emph{OpenAI GPT} heading for readability, as they share the same provider. We do not, however, treat them as interchangeable: a calibration study (\S\ref{sec:setup-calibration}) shows that the two models differ in the characteristics of the generated code. Accordingly, all values reported under this heading are interpreted as specific to the model used in each language (ChatGPT for Python and Java, gpt-oss for C), and we draw no cross-language conclusions that depend on the two models being equivalent.
 
To generate the LLM-implemented version of each sample, we followed the default generation settings recommended by each model's authors and used their preferred prompting formats, with a maximum output length of 512 tokens. This budget comfortably exceeds the average length of the human-written functions across all three languages (Table~\ref{tab:dataset_stats}), accommodating the typical generation while keeping inference cost tractable across the full corpus. Example prompts for each LLM for Python, Java, and C generation are shown in the listings below.
 
\begin{mainbox}{}
\label{AI_prompts}
\footnotesize
You are an AI programming assistant, utilizing the DeepSeek-Coder model, developed by DeepSeek Company, and you only answer questions related to computer science.\\
\textbf{\#\#\# Instruction:}\\
Java\\
\texttt{"""\{docstring\}"""}\\
\texttt{\{signature\}}\\
\textbf{\#\#\# Response:}
\end{mainbox}

\begin{mainbox}{}
\footnotesize
You are Qwen, created by Alibaba Cloud. You are a helpful assistant.\\
Python\\
\texttt{\{signature\}}\\
\texttt{"""\{docstring\}"""}
\end{mainbox}

\begin{mainbox}{}
\footnotesize
You are a careful C code generation assistant. Use the exact function signature provided. Complete only this function.\\
\textbf{User:} Implement the following C function from its docstring.\\
Docstring:\\
\texttt{\{docstring\}}\\
Signature:\\
\texttt{\{signature\}}
\end{mainbox}

Following generation, a comprehensive code cleanup phase was applied to ensure consistency and validity across all samples. The cleanup process included normalization of indentation, removal of leading or trailing whitespace inconsistencies, and lightweight syntactic parsing checks to detect incomplete or invalid code fragments. In addition, special care was taken to remove any non-code text artifacts often produced by AI models, such as explanations of the code, usage examples, comments unrelated to the function's purpose, or generic model disclaimers. To ensure the preservation of valid code, the cleaning procedure was refined iteratively: several rounds of manual inspection were conducted on random subsets of the data to verify that correct code instances were not inadvertently discarded or damaged. Cleaning rules were adjusted as necessary to minimize the loss of properly generated samples.

The final dataset comprises \textbf{787,562} $\langle$\emph{docstring}, \emph{human-code}, \emph{OpenAI-code}, \emph{DSC-code}, \emph{Qwen-code}$\rangle$ tuples across the three languages: 285,249 for Python, 221,795 for Java, and 280,518 for C, and it's available in the replication package~\cite{replication}.

\subsection{Model consistency calibration}
\label{sec:setup-calibration}
 
Because the \emph{OpenAI GPT} category comprises ChatGPT (``gpt-3.5-turbo'') for Python and Java and gpt-oss-20B for C, we assess whether the two models behave equivalently before reporting them under a common heading. We generated a gpt-oss-20B implementation for 1,000 randomly selected Python and 1,000 Java specifications already implemented by ChatGPT in HMCorp, and applied the analysis procedures of \S\ref{sec:research_study} to both. We compare the row-matched pairs with paired Wilcoxon signed-rank tests and report Cliff's $\delta$ as effect size, applying Benjamini-Hochberg correction across metrics. The complete per-metric panel is available in the replication package~\cite{replication}.
 
The two models are not directly interchangeable. gpt-oss generates longer functions than ChatGPT in both languages, while defect and vulnerability results are language-dependent. In Python, gpt-oss triggers more Pylint findings than ChatGPT ($p<0{.}001$) but fewer security vulnerable functions. In Java, the two models produce comparable defect \emph{volumes} (no significant difference in total findings), although their distribution across ODC categories shifts slightly, yet gpt-oss generates less vulnerable but also more error-prone code.
 
These differences concern the \emph{magnitude} of the measurements, not the comparative structure of the study. The relative ordering of code authors is preserved under the substitution: in both models the OpenAI-family entry is the most \emph{human-like} of the LLM authors, and the qualitative ranking among human, OpenAI, DeepSeek, and Qwen is unchanged. We therefore retain \emph{OpenAI GPT} as a provider-level reporting label, but interpret all values as specific to the model used in each language, and we draw no cross-language conclusions that depend on the two models being equivalent. The absolute human-vs-AI complexity gap under this heading is consequently model-dependent and is read per language throughout \S\ref{sec:evaluation}.

\section{Experimental Setup}
\label{sec:setup}
\paragraph{\textbf{Code generation capabilities.}}
The four AI coding assistants we employ are established, state-of-the-art code generators, selected for their prominence in recent code-generation benchmarks and their active use in development workflows. On the widely used HumanEval and MBPP Python benchmarks, DeepSeek-Coder-33B-Instruct reports a pass@1 of roughly 79\%, also exceeding gpt-3.5-turbo on HumanEval~\cite{guo2024deepseek}, while Qwen2.5-Coder-32B-Instruct reports pass@1 around 88--92\% on HumanEval and 84--90\% on MBPP~\cite{hui2024qwen2}. ChatGPT (gpt-3.5-turbo), the OpenAI model used for Python and Java, performs in a comparable range on these Python benchmarks. Beyond Python, both open models report strong multilingual results on MultiPL-E (average pass@1 of about 68\% for DeepSeek-Coder and 75\% for Qwen2.5-Coder, including C++ and Java)~\cite{guo2024deepseek,hui2024qwen2}, and other work performing multi-language evaluations places them in the same tier on C++ specifically~\cite{cheng2024fullstack}.

We further note that public benchmarks for C function generation are scarce: the standard multilingual suites (MultiPL-E, HumanEval-X) cover C++ rather than C, and OpenAI evaluates gpt-oss-20B on agentic coding benchmarks such as Codeforces and SWE-bench Verified rather than pass@1 function synthesis~\cite{gptoss}. Current literature places gpt-oss-20B in the mid tier of current open models, reaching 73\% pass@1 on HumanEval, with code generation noted as a relative strength of the model at its parameter scale~\cite{bi2025gpt}. Because no benchmark isolates the C generation quality of these models, our calibration (\S\ref{sec:setup-calibration}) and analyses provide the relevant evidence of their behavior on our task.

\paragraph{\textbf{Code generation infrastructure.}}
We generated the LLM-implemented version of each sample from the original docstring and the function signature extracted from the human code, following the prompting scheme detailed in \S\ref{sec:dataset}. DeepSeek-Coder and Qwen were served with vLLM~\cite{kwon2023efficientmemorymanagementlarge} across all three languages, while gpt-oss was generated through a Hugging Face Transformers pipeline with Harmony-format chat templating. Experiments were performed on four NVIDIA A100 (40~GB) GPUs.

\paragraph{\textbf{Structural Complexity Computation.}}
\label{sec:setup-complexity}

We compute the structural complexity metrics from \S\ref{sec:complexity} using \textit{lizard}~\cite{lizard} for function extraction, NLOC, Cyclomatic Complexity Number, and Parameter Count, and a custom lexical pipeline for the remaining measures. Halstead metrics and Maximum Nesting Depth are computed from per-language token streams produced with Pygments. Tokens are classified as operators (keywords, operators, and punctuation) or operands (identifiers and literals), yielding $\eta_1$, $N_1$, $\eta_2$, and $N_2$, from which we derive Halstead Volume, Difficulty, and Effort~\cite{halstead1977elements}. Nesting depth is computed over the same stream by tracking control-flow block entry and exit, with continuation headers such as \texttt{else}, \texttt{finally}, and \texttt{case} attached to their enclosing construct.

Maintainability Index follows the Microsoft-adjusted Oman--Hagemeister formulation~\cite{oman1992mi}. Function Name Length is the character count of the name returned by lizard, and corpus-level Unique Tokens is the number of distinct lexical units across an author's corpus. Metrics are collected separately per author; all scripts are released in the replication package~\cite{replication}.

\paragraph{\textbf{Code Naturalness Computation.}}
\label{sec:setup-naturalness}

We train $n$-gram language models with KenLM~\cite{heafield2011kenlm}, which supports efficient entropy estimation at the million-function scale and is widely used in software-naturalness, security-auditing, and attack-detection studies~\cite{liu2018s,gholamian2021naturalness,sun2025show}. Comments and docstrings are removed before training so that the models capture code style and syntax rather than documentation. For each author, samples are concatenated into one file with one function per line and tokenized using a deterministic, language-agnostic tokenizer to ensure comparability across languages and authors while avoiding parser failures on partial or syntactically irregular generated code. The tokenizer discards whitespace, groups consecutive word characters into single tokens, and emits every other non-whitespace symbol individually. This preserves identifiers and literals as atomic lexical units while isolating operators and punctuation, which are important carriers of code structure. We intentionally avoid language-specific normalization at this stage; lexical abstraction is evaluated separately through the identifier-normalized setting defined in \S\ref{sec:naturalness-design}.

For each author and language, we train models with $n \in [2,10]$ under 10-fold cross-validation, using 90\% of the author’s code for training and 10\% for testing in each fold. Since self-cross-entropy plateaus at $n=6$ for all three languages, we use 6-grams for the cross-author transferability and clustering analyses. The identifier-normalization scheme used to separate lexical from structural contributions to naturalness is defined in \S\ref{sec:naturalness-design}. Cross-entropy is reported in bits per token and perplexity as its exponentiation. Training and scoring scripts are included in the replication package~\cite{replication}.

\paragraph{\textbf{Defect Analysis Configuration.}}
\label{sec:setup-defects}

We analyze defects with Pylint~\cite{pylint} for Python, PMD~\cite{pmd} for Java, and Clang-Tidy~\cite{clang} for C. As described in \S\ref{sec:research_study}, we exclude style, documentation, and security rules, the latter being handled separately. The retained configuration contains 350 Python rules, 226 Java rules, and 258 C rules targeting functional, structural, or design defects at the function level. The tool version, full retained/excluded rule list and rule-to-ODC mapping are provided in the replication package~\cite{replication}.

Two adjustments keep the analysis function-scoped. First, because PMD requires valid Java class structures, each Java function is wrapped in a minimal dummy class. To avoid artifacts from this synthetic context, we exclude 12 additional class-level PMD rules, including \textit{UseUtilityClass} and \textit{ClassWithOnlyPrivateConstructorsShouldBeFinal}. Second, for Clang-Tidy, we remove checks targeting features outside our C setting, such as C++, Objective-C, OS-X, and WebKit-specific rules. These checks are nevertheless pre-classified in the mapping to support future extensions to C++ and Objective-C.

\paragraph{\textbf{Security Analysis Configuration.}}
\label{sec:setup-security}

We perform static security analysis with Semgrep OSS~\cite{semgrep}, applying its curated security-focused rule sets uniformly across the three languages. We include all available security-focused rules for each language. For each finding, Semgrep reports the triggering rule, severity level, source lines, human-readable weakness description, and associated Common Weakness Enumeration identifier. We use the CWE metadata as provided, without manual reclassification. The complete security configuration and rule list are included in the replication package~\cite{replication}.

\paragraph{\textbf{Statistical Analysis.}}
\label{sec:setup-statistics}

Comparisons between code authors are performed on row-matched function pairs: each human function is compared with the AI implementation generated from the same docstring and signature. We test paired differences using the Wilcoxon signed-rank test and control the false discovery rate across metrics with the Benjamini--Hochberg correction. As effect-size measures, we report the median paired difference with 95\% bootstrap confidence intervals computed over 1,000 paired resamples.

\section{Results}
\label{sec:evaluation}
This section illustrates the results of our investigation. 
First, we examine whether the structure, complexity, and style of code differ between human developers and AI code generators, characterizing each author through function-level complexity metrics and the statistical naturalness of the code across Python, Java, and C (\S\ref{sec:results-complexity}). Then, we dive into the defect categorization and distribution between human-written and AI-generated code: we show which ODC defect types occur most and the frequency with which each specific code issue appears in each language (\S\ref{sec:results-defects}). Next, we assess the security vulnerabilities introduced by human developers and AI code generators, respectively, classifying them according to MITRE's CWE and analyzing their distribution and severity (\S\ref{sec:results-vulns}). Finally, we examine whether the structural and statistical properties of the code are associated with its defect and vulnerability profiles (\S\ref{sec:results-correlations}).

\subsection{RQ$_1$: Do code structural properties and style differ between human-written and AI-generated code across programming languages?}
\label{sec:results-complexity}

\begin{table*}[t]
  \caption{Comparison of structural complexity metrics across languages and code authors. Except for UT, each author row reports the mean $\pm$ standard deviation across samples. UT is the corpus-level unique-token vocabulary and is reported as a single value per author. The LLM-avg row is the arithmetic mean of the three model-specific values. \textcolor{blue}{\textbf{Blue bold}} values indicate the highest author mean or UT value for each language. Five additional metrics ($\eta_1$, $N_1$, $\eta_2$, $N_2$, $E$) follow the same pattern and are reported in the replication package~\cite{replication}.}
  \label{tab:complexity-summary}
  \centering
  \scriptsize
  \setlength{\tabcolsep}{4pt}
  \begin{tabular}{llrrrrrrrrr}
  \toprule
  Language & Author & NLOC & CCN & PC & MND & $V$ & $D$ & MI & FNL & UT \\
  \midrule

  \multirow{5}{*}{Python}
   & Human
   & \textcolor{blue}{\textbf{12.76}} $\pm$ 15.37
   & \textcolor{blue}{\textbf{4.02}} $\pm$ 4.61
   & \textcolor{blue}{\textbf{2.32}} $\pm$ 1.54
   & \textcolor{blue}{\textbf{1.25}} $\pm$ 1.10
   & \textcolor{blue}{\textbf{611.40}} $\pm$ 1,019.26
   & \textcolor{blue}{\textbf{16.66}} $\pm$ 10.61
   & 60.56 $\pm$ 10.26
   & 13.65 $\pm$ 7.12
   & \textcolor{blue}{\textbf{70,542}} \\

   & \cellcolor{gray!15}\textit{LLM-avg}
   & \cellcolor{gray!15}\textit{6.56 $\pm$ 5.14}
   & \cellcolor{gray!15}\textit{2.47 $\pm$ 2.01}
   & \cellcolor{gray!15}\textit{1.97 $\pm$ 1.19}
   & \cellcolor{gray!15}\textit{0.86 $\pm$ 0.93}
   & \cellcolor{gray!15}\textit{273.75 $\pm$ 298.87}
   & \cellcolor{gray!15}\textit{11.88 $\pm$ 7.37}
   & \cellcolor{gray!15}\textit{68.46 $\pm$ 9.02}
   & \cellcolor{gray!15}\textit{14.82 $\pm$ 6.57}
   & \cellcolor{gray!15}\textit{62,060} \\

   & OpenAI GPT
   & 6.80 $\pm$ 5.75
   & 2.52 $\pm$ 2.15
   & 1.66 $\pm$ 1.15
   & 0.90 $\pm$ 1.01
   & 287.74 $\pm$ 380.88
   & 12.24 $\pm$ 8.03
   & 67.96 $\pm$ 9.20
   & \textcolor{blue}{\textbf{17.57}} $\pm$ 6.44
   & 62,717 \\

   & DeepSeek
   & 5.66 $\pm$ 3.50
   & 2.18 $\pm$ 1.47
   & 2.08 $\pm$ 1.16
   & 0.77 $\pm$ 0.85
   & 214.97 $\pm$ 164.93
   & 10.86 $\pm$ 5.79
   & \textcolor{blue}{\textbf{69.85}} $\pm$ 8.13
   & 13.15 $\pm$ 6.46
   & 60,199 \\

   & Qwen
   & 7.23 $\pm$ 6.18
   & 2.71 $\pm$ 2.42
   & 2.16 $\pm$ 1.26
   & 0.91 $\pm$ 0.94
   & 318.54 $\pm$ 350.80
   & 12.55 $\pm$ 8.29
   & 67.57 $\pm$ 9.72
   & 13.76 $\pm$ 6.80
   & 63,265 \\

  \midrule

  \multirow{5}{*}{Java}
   & Human
   & \textcolor{blue}{\textbf{14.09}} $\pm$ 18.61
   & \textcolor{blue}{\textbf{3.68}} $\pm$ 5.74
   & \textcolor{blue}{\textbf{1.48}} $\pm$ 1.16
   & \textcolor{blue}{\textbf{1.08}} $\pm$ 1.15
   & \textcolor{blue}{\textbf{601.78}} $\pm$ 1,178.43
   & \textcolor{blue}{\textbf{16.50}} $\pm$ 10.60
   & 59.78 $\pm$ 10.26
   & 13.78 $\pm$ 7.62
   & \textcolor{blue}{\textbf{57,356}} \\

   & \cellcolor{gray!15}\textit{LLM-avg}
   & \cellcolor{gray!15}\textit{7.80 $\pm$ 5.84}
   & \cellcolor{gray!15}\textit{2.22 $\pm$ 1.91}
   & \cellcolor{gray!15}\textit{1.43 $\pm$ 1.03}
   & \cellcolor{gray!15}\textit{0.75 $\pm$ 0.88}
   & \cellcolor{gray!15}\textit{296.50 $\pm$ 293.40}
   & \cellcolor{gray!15}\textit{12.42 $\pm$ 7.36}
   & \cellcolor{gray!15}\textit{66.66 $\pm$ 9.07}
   & \cellcolor{gray!15}\textit{16.73 $\pm$ 8.27}
   & \cellcolor{gray!15}\textit{50,325} \\

   & OpenAI GPT
   & 8.95 $\pm$ 7.18
   & 2.51 $\pm$ 2.43
   & 1.48 $\pm$ 1.01
   & 0.87 $\pm$ 0.96
   & 352.19 $\pm$ 383.82
   & 13.89 $\pm$ 8.54
   & 64.53 $\pm$ 8.91
   & \textcolor{blue}{\textbf{18.37}} $\pm$ 8.02
   & 52,540 \\

   & DeepSeek
   & 7.61 $\pm$ 4.32
   & 2.06 $\pm$ 1.41
   & 1.36 $\pm$ 0.96
   & 0.74 $\pm$ 0.79
   & 279.72 $\pm$ 207.16
   & 12.40 $\pm$ 5.93
   & 65.98 $\pm$ 7.65
   & 18.11 $\pm$ 9.31
   & 47,737 \\

   & Qwen
   & 6.84 $\pm$ 6.02
   & 2.10 $\pm$ 1.89
   & 1.46 $\pm$ 1.13
   & 0.65 $\pm$ 0.88
   & 257.60 $\pm$ 289.21
   & 10.96 $\pm$ 7.60
   & \textcolor{blue}{\textbf{69.47}} $\pm$ 10.66
   & 13.72 $\pm$ 7.48
   & 50,698 \\

  \midrule

  \multirow{5}{*}{C}
   & Human
   & \textcolor{blue}{\textbf{26.41}} $\pm$ 25.43
   & \textcolor{blue}{\textbf{5.97}} $\pm$ 6.90
   & \textcolor{blue}{\textbf{2.17}} $\pm$ 1.72
   & \textcolor{blue}{\textbf{1.13}} $\pm$ 1.12
   & \textcolor{blue}{\textbf{1,030.09}} $\pm$ 1,180.52
   & \textcolor{blue}{\textbf{21.79}} $\pm$ 14.30
   & 52.08 $\pm$ 11.67
   & \textcolor{blue}{\textbf{18.33}} $\pm$ 7.85
   & \textcolor{blue}{\textbf{60,120}} \\

   & \cellcolor{gray!15}\textit{LLM-avg}
   & \cellcolor{gray!15}\textit{11.36 $\pm$ 7.56}
   & \cellcolor{gray!15}\textit{3.29 $\pm$ 2.45}
   & \cellcolor{gray!15}\textit{2.00 $\pm$ 1.57}
   & \cellcolor{gray!15}\textit{0.85 $\pm$ 0.83}
   & \cellcolor{gray!15}\textit{400.32 $\pm$ 325.71}
   & \cellcolor{gray!15}\textit{15.57 $\pm$ 9.27}
   & \cellcolor{gray!15}\textit{62.23 $\pm$ 9.26}
   & \cellcolor{gray!15}\textit{17.56 $\pm$ 7.84}
   & \cellcolor{gray!15}\textit{51,388} \\

   & OpenAI GPT
   & 14.51 $\pm$ 9.54
   & 4.68 $\pm$ 3.52
   & 2.14 $\pm$ 1.66
   & 0.93 $\pm$ 0.91
   & 553.76 $\pm$ 447.97
   & 19.94 $\pm$ 11.05
   & 57.67 $\pm$ 8.85
   & 18.31 $\pm$ 7.84
   & 53,514 \\

   & DeepSeek
   & 7.05 $\pm$ 4.61
   & 1.99 $\pm$ 1.34
   & 1.75 $\pm$ 1.44
   & 0.61 $\pm$ 0.70
   & 216.54 $\pm$ 181.42
   & 10.92 $\pm$ 7.26
   & \textcolor{blue}{\textbf{68.15}} $\pm$ 8.80
   & 16.25 $\pm$ 7.95
   & 50,195 \\

   & Qwen
   & 12.51 $\pm$ 8.52
   & 3.21 $\pm$ 2.48
   & 2.10 $\pm$ 1.62
   & 1.02 $\pm$ 0.89
   & 430.65 $\pm$ 347.73
   & 15.85 $\pm$ 9.49
   & 60.89 $\pm$ 10.14
   & 18.12 $\pm$ 7.74
   & 50,454 \\

  \bottomrule
  \end{tabular}
  \end{table*}

\subsubsection{Structural complexity}
\label{sec:complexity}

We measure structural complexity for each function with the metric suite defined in \S~\ref{sec:complexity-design}. \tableautorefname~\ref{tab:complexity-summary} summarizes the nine metrics retained in the main results across all four code authors per language, with an additional \textit{LLM-avg} row indicating the simple mean over the three AI coding assistants. Five additional metrics ($\eta_1$, $N_1$, $\eta_2$, $N_2$, $E$) follow the same pattern and are reported in the replication package~\cite{replication}.
Paired Wilcoxon signed-rank tests on row-matched function pairs confirm the directions of all cross-author differences reported in Table~\ref{tab:complexity-summary}, all significant at $p < 0.001$ after Benjamini-Hochberg correction.

On size, human-written code is the structural outlier in every language. The mean NLOC, \ie number of lines of code, of human functions is 12.76 in Python, 14.09 in Java, and 26.41 in C, against an \textit{LLM-avg} of 6.56, 7.80, and 11.36 respectively, with ratios of 1.95$\times$, 1.81$\times$, and 2.32$\times$. The gap widens in C, where the human surplus is largest in both relative and absolute terms. Dispersion follows the same shape: the human NLOC standard deviation is roughly three times the \textit{LLM-avg} in every language (Python 15.37 vs 5.14; Java 18.61 vs 5.84; C 25.43 vs 7.56), reflecting that humans span a wider distribution of function sizes than any LLM produces.

On control-flow complexity, the profile mirrors the size profile but with a sharper edge in C than in Python and Java. Human is the most branching by CCN, \ie cyclomatic complexity number, (4.02 / 3.68 / 5.97) and the deepest nesting by MND, \ie max nesting depth, (1.25 / 1.08 / 1.13); \textit{LLM-avg} ratios are 1.63$\times$, 1.66$\times$, and 1.81$\times$ on CCN and 1.45$\times$, 1.44$\times$, and 1.33$\times$ on MND.
Parameter counts follow this trend but compress the signal.
The C-specific pattern is notable: C gpt-oss reaches CCN 4.68 and MND 0.93, the only LLM in any language to approach the human range on branching depth, while DeepSeek collapses to CCN 1.99 and MND 0.61 and Qwen to 3.21 and 1.02. The structural simplification of LLM-generated code is therefore not a uniform effect: it is mediated by the specific model.

On lexical and data complexity, the Halstead metrics extend the size and branching story onto the data dimension. Human dominates volume $V$ at 2.23$\times$, 2.03$\times$, and 2.57$\times$ over \textit{LLM-avg}, and dominates difficulty $D$ at 1.40$\times$, 1.33$\times$, and 1.40$\times$. Taken together with the size and control-flow categories, these numbers describe a code surface that is uniformly larger, more branched, and more operator- and operand-dense in human-written code than in any LLM output, in every language. The five additional Halstead metrics reported in the appendix ($\eta_1$, $N_1$, $\eta_2$, $N_2$, $E$) reinforce the same ordering.
 
On corpus-level lexical diversity, human software reaches 70,542 unique tokens in Python, 57,356 in Java, and 60,120 in C; the \textit{LLM-avg} sits at 62,060, 50,325, and 51,388, placing the LLMs at 88\%, 88\%, and 85\% of the human vocabulary respectively. The LLMs therefore command a substantial fraction of the lexical space of their human counterparts, but reuse that vocabulary across functions at a noticeably higher rate, which may be a symptom of more templated phrasing. This reading is consistent with the structural compression observed for other metrics: when an AI coding assistant writes shorter, less branching functions, the chance of reusing the same set of tokens increases.
 
Regarding the maintainability of code, the pattern from every previous category inverts. Every LLM exceeds the human MI baseline in every language, with \textit{LLM-avg} surpluses of 7.90 (Python), 6.88 (Java), and 10.15 (C) absolute MI points, the largest reversal in the entire table. We treat this inversion with caution rather than as a finding. MI is a deterministic function of NLOC, Halstead Volume, and CCN with negative coefficients on each, so any author that produces consistently smaller, less branching functions will mechanically achieve a higher MI regardless of the actual quality of the code. The MI ``win'' for LLMs may therefore be a consequence of their smaller NLOC and $V$, not an independent quality signal. The second metric in this category, FNL, tells a more interesting story: in Python and Java the LLMs exceed the human baseline (OpenAI GPT 17.57 in Python vs human 13.65; OpenAI GPT 18.37 in Java vs human 13.78), while in C the human baseline of 18.33 is the highest of any author. The C reversal reflects the language ecosystem: human-written C software is drawn primarily from systems and kernel code where descriptive function names are already the convention, so the LLMs have less space to outdo their training data on identifier verbosity.
 
The five dimensions describe a coherent picture, but not the one suggested by the MI column. On size, control flow, and Halstead complexity, AI-generated code is uniformly compressed: roughly half the NLOC, well below the human branching, and one third to one half the Halstead Volume of human-written code. The models differ among themselves in the degree of compression, and the ordering is language-dependent: DeepSeek is the most compressed LLM in every language, while the least-compressed, most human-like LLM is OpenAI GPT in Java and C but Qwen in Python. The lexical diversity category shows that this compression is also reflected in vocabulary reuse. LLM outputs are smaller likely because they implement narrower behavior, omit defensive code paths, and emit templated patterns drawn from training data: the size, branching, and Halstead surpluses of human-written code reflect functionality the LLMs systematically leave out, including error handling, edge cases, configuration logic, and parameter validation, rather than complexity.

\subsubsection{Code naturalness}

\begin{figure}[h!]
\centering
\includegraphics[width=\linewidth]{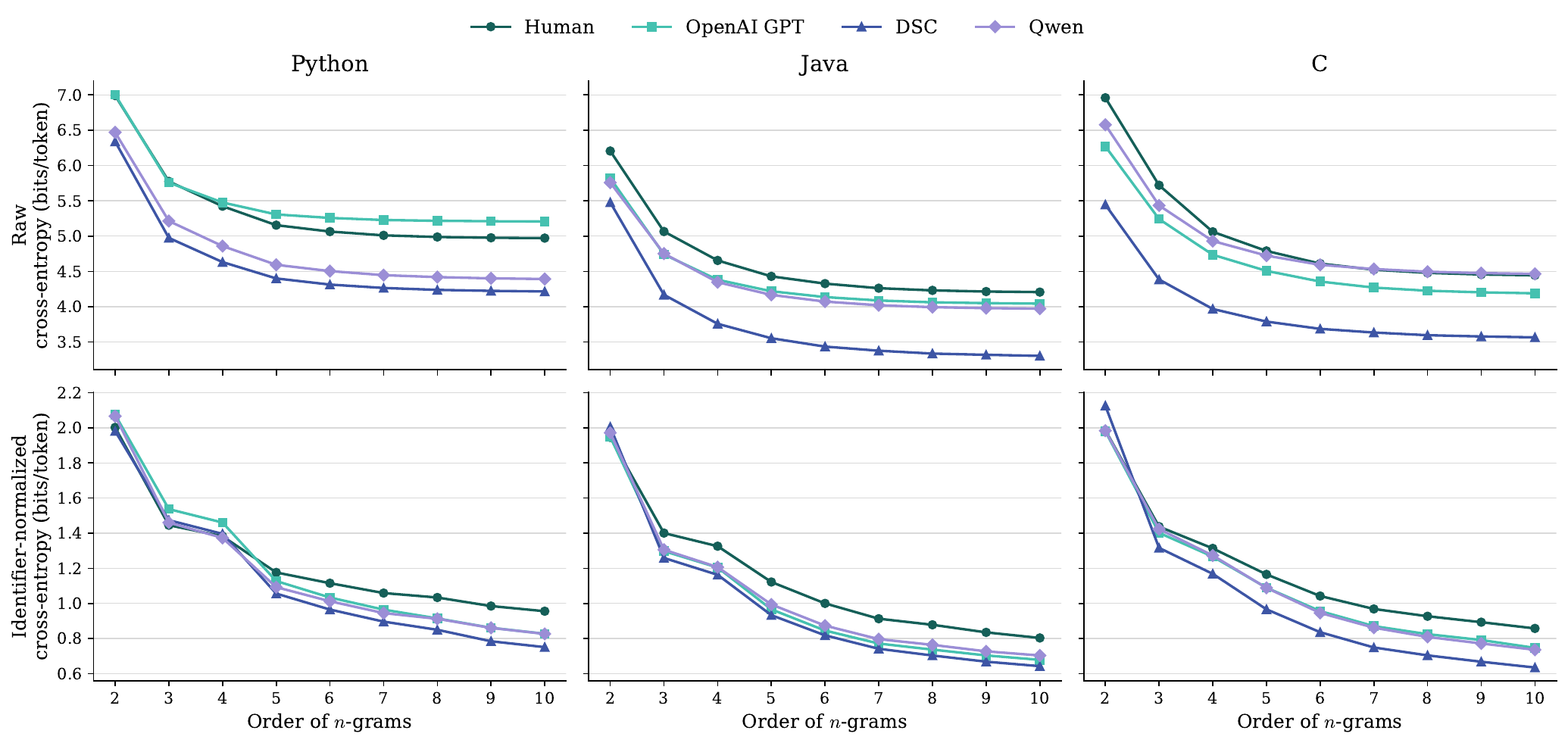}
\caption{Self-cross-entropy of each author's code under its own $n$-gram model,
10-fold cross-validated, for the raw token stream (top row) and the
identifier-normalized representation (bottom
row). Raw curves flatten around $n=6$, which fixes the order used in
Figures~\ref{fig:perplexity} and~\ref{fig:dendrogram}; normalized curves continue
to decline gently, so normalized comparisons are read at $n=10$. In C, OpenAI GPT denotes gpt-oss-20B.}
\label{fig:selfcrossentropy}
\end{figure}

\begin{figure}[h!]
\centering
\includegraphics[width=\linewidth]{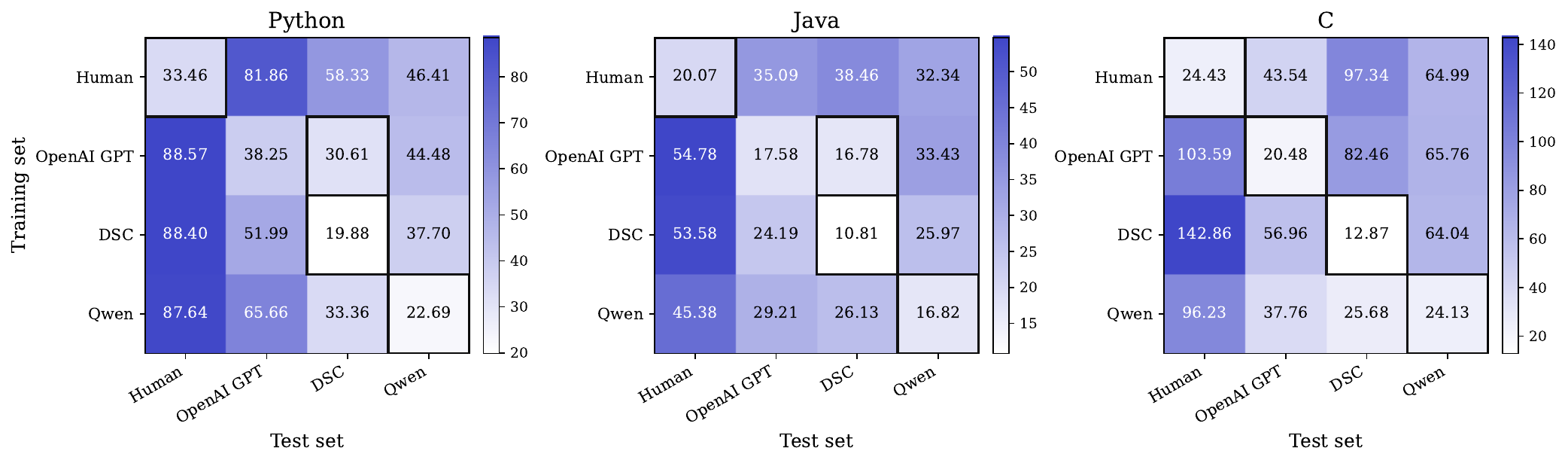}
\caption{Cross-perplexity at $n=6$: each row is an $n$-gram model trained on one author's code, each column a held-out test corpus; each cell reports $\mathrm{PP}_{X \to Y} = 2^{H_{X \to Y}}$. Bordered cells mark each row's minimum.}
\label{fig:perplexity}
\end{figure}

\begin{figure}[h!]
\centering
\includegraphics[width=\linewidth]{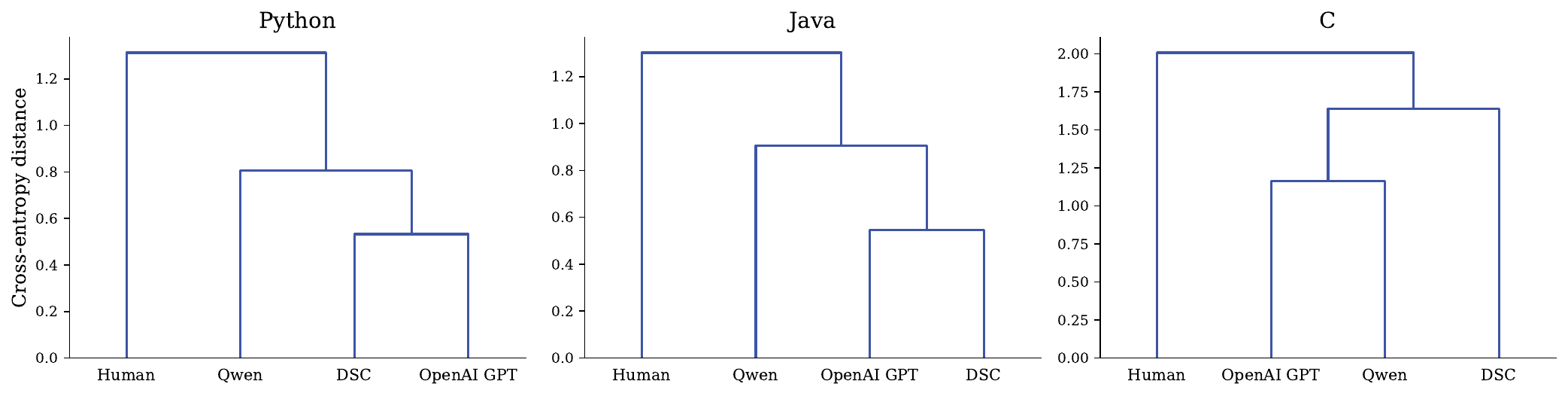}
\caption{Average-linkage agglomerative clustering of authors on the symmetrized excess cross-entropy distance of Equation~\ref{eq:excess_ce_distance} at $n=6$.}
\label{fig:dendrogram}
\end{figure}

We assess code naturalness with n-gram language models following the methodology of \S~\ref{sec:naturalness-design}, performing the four complementary analyses defined there: intrinsic naturalness via self-cross-entropy, the contribution of naming versus structural syntax via identifier-normalized self-cross-entropy, cross-author transferability via the cross-perplexity matrix, and stylistic affinity via hierarchical clustering. KenLM n-gram models are trained for each author and each language with $n$ ranging from 2 to 10 under 10-fold cross-validation, and the saturation point is fixed at $n=6$ where the raw self-cross-entropy curves plateau in all three languages; the normalized curves of the second analysis decline gently through $n=10$ and are read there.

Intrinsic naturalness, \ie how locally regular and predictable each author's code is on its own~\cite{hindle2016naturalness}, is captured by the self-cross-entropy curves in the top row of \figureautorefname~\ref{fig:selfcrossentropy}. All four curves decrease monotonically with $n$ and flatten around $n=6$, indicating that most author-specific regularity is captured by short to mid-range context windows. DSC has the lowest self-cross-entropy at the plateau in every language by a substantial margin (Python 4.31; Java 3.43; C 3.69), confirming that its code is the most locally regular and templated of the four authors. The ordering of the remaining authors is language-dependent in ways that resist a uniform ``\textit{LLMs are more predictable than human}'' reading. In Java all three LLMs sit below the human baseline of 4.33. In Python, ChatGPT sits above human at the plateau (5.26 vs 5.06), making ChatGPT's Python code the \textit{least} predictable of the four. In C, Qwen is essentially tied with human (4.59 vs 4.61). At the surface, the relationship between authorship and intrinsic naturalness is therefore not uniform: it depends on the specific LLM's stylistic conventions.

Naming and structural syntax contribute jointly to self-cross-entropy, and the unnormalized curves cannot tell them apart: an n-gram model treats identifier strings and structural tokens as units of the same alphabet, so an author whose code repeats the same identifier patterns and an author whose code repeats the same syntactic patterns both register as locally predictable. To account for these differences, we repeat the self-cross-entropy procedure on a normalized version of each corpus where every identifier and literal is replaced with a role-aware placeholder tagging its syntactic role (functions, classes, parameters, variables, attributes, modules, types, and literal categories; see \S~\ref{sec:naturalness-design}), while all structural tokens (keywords, delimiters, operators, parentheses, braces, and control-flow constructs) are preserved verbatim. 
This transformation collapses the otherwise large and sparse identifier vocabulary into a compact set of category tokens, retaining the syntactic and structural form of each program but removing naming-level variation; the residual self-cross-entropy then reflects predictability driven by structural patterns alone, and the difference between the original and normalized curves quantifies the lexical contribution. 

The resulting normalized curves are shown in the bottom row of \figureautorefname~\ref{fig:selfcrossentropy}, and they resolve the surface inconsistencies observed above. The relative ordering of authors clarifies: in every language, human-authored code now sits at the highest cross-entropy at the plateau, with the three LLMs uniformly below. The inversion on Python disappears: at $n=10$ the normalized human curve is at roughly 0.95 bits per token and OpenAI GPT at roughly 0.82, with Qwen at 0.82 and DSC at 0.75. The C result similarly resolves into Qwen at roughly 0.75 against human at 0.85. The two language-specific anomalies of the unnormalized curves are therefore naming artifacts: OpenAI GPT's varied Python identifiers and Qwen's relatively diverse C identifiers inflate the surface entropy above the underlying structural patterns. The normalized residual establishes the harder claim. Even after the bulk of the gap is attributed to naming, a consistent structural difference between human and LLM code persists across all three languages and is robust to the specific LLM. The decomposition is therefore not ``the gap is purely lexical'' but ``naming variation is the larger component, and a structural component remains.''

Cross-author transferability quantifies whether an author's stylistic regularities are specific to that author or shared more broadly. \figureautorefname~\ref{fig:perplexity} reports the cross-perplexity matrix at $n=6$ for each language, with each row corresponding to an n-gram model trained on a single author and each column to a held-out test corpus. The diagonal is the lowest cell of each row in C and in nearly every row in Python and Java (bordered cells in \figureautorefname~\ref{fig:perplexity}), confirming that each author's n-gram model best predicts that author's own code, but the off-diagonal structure reveals a marked human-LLM asymmetry. Models trained on LLM corpora struggle to predict human code, with LLM-trained perplexity on human ranging from 88 to 89 in Python, 45 to 55 in Java, and 96 to 143 in C. 

Models trained on human code predict LLM corpora with substantially lower penalty. Human-written code is therefore the hardest target for n-gram models trained on machine output, while machine output is comparatively predictable from a human-trained model. Within the LLMs, the Python and Java matrices show that OpenAI GPT and DeepSeek are mutually predictive to a degree that breaks the diagonal rule: OpenAI GPT-trained models predict the DeepSeek test set more easily than themselves (Python 30.61 vs 38.25; Java 16.78 vs 17.58), the only LLM-vs-LLM cells in either language where the diagonal is not the minimum. The C matrix does not reproduce this coupling. %DeepSeek-C has the lowest self-perplexity in the study (12,87), but its cross-perplexity to gpt-oss (56,96), Qwen (64,04), and human (142,86) are uniformly higher than the corresponding Python and Java cells, indicating that DeepSeek-C occupies a stylistic region not shared by the other two C LLMs and that the within-LLM cohesion observed in Python and Java does not generalize to C.

Stylistic affinity at the author level is captured by average-linkage hierarchical clustering on the excess cross-entropy distance of Equation~\ref{eq:excess_ce_distance} at $n=6$, shown in \figureautorefname~\ref{fig:dendrogram}. The top-level split in every language separates human from the LLM cluster, confirming that AI-generated code is statistically closer to other AI-generated code than to human-written code regardless of the specific model. Within the LLM cluster, the internal organization varies by language. In Python and Java the dendrograms place Qwen adjacent to the human branch and pair OpenAI GPT and DeepSeek as nearest neighbors. In C the role of the human-adjacent LLM shifts from Qwen to OpenAI GPT, and DeepSeek attaches at the greatest distance from human; the C dendrogram also extends to a larger absolute distance scale, around 2.0 against 1.3 in Python and Java, indicating greater inter-author dispersion in the C corpus. The C results are consistent with the structural complexity findings in \S~\ref{sec:complexity}, where OpenAI GPT is the only LLM to approach human values on size, branching, and Halstead complexity.

The four analyses converge with the structural complexity findings of \S~\ref{sec:complexity}. Surface self-cross-entropy conflates structural and lexical variation, and in some cases (OpenAI GPT in Python, Qwen in C) the LLM's verbose or diverse naming inflates the surface entropy above the human baseline. The normalization analysis decomposes the signal: most of the human-LLM gap is in identifier choice, but a residual structural difference holds in every language with human at the top and DeepSeek at the bottom. The cross-perplexity matrices and the dendrograms confirm the same partition from a different angle, with machine-generated code clustering together and statistically further from human code than from itself. AI-generated code is therefore both structurally compressed at the function level, as documented in \S~\ref{sec:complexity}, and more templated at the token-sequence level, as documented here; the maintainability advantage observed on MI can be interpreted as a consequence of the structural compression rather than of independently better engineering, and the apparent surface diversity of certain LLM outputs is identifier-driven rather than evidence of richer underlying program structure.

\begin{mainbox}{}
AI-generated code differs from human-written code on two coupled axes: structural compression and templated lexicon. Every LLM produces functions with roughly half the NLOC, half the cyclomatic complexity, and one third to one half the Halstead Volume of human code in every language. After identifier and literal masking, a uniform naturalness ordering emerges with human at the highest cross-entropy and DeepSeek-Coder at the lowest, indicating a structural gap masked by lexical variation at the surface; cross-perplexity matrices and dendrograms confirm the same partition. DSC is the extreme case on both axes in every language, while OpenAI GPT is the closest LLM to human on structural complexity in C and the LLM nearest to human in the C dendrogram.
\end{mainbox}

\subsection{RQ$_2$: Do defect types and frequencies differ between human-written and AI-generated code across programming languages?}
\label{sec:results-defects}

To investigate differences in the nature and frequency of code defects across human-written and AI-generated code, we analyze each code sample (Pylint for Python, PMD for Java, Clang-Tidy for C), and categorize the resulting violations using the Orthogonal Defect Classification framework to ensure consistency across languages and tools. 
We compare the four code authors per language: Human, OpenAI GPT (ChatGPT for Python and Java; gpt-oss for C), DeepSeek-Coder (DSC), and Qwen2.5-Coder (Qwen). 
To increase interpretability of the results, this section provides an overview of our findings; complete per-author and per-rule defect distributions are available in the replication package~\cite{replication}.

\begin{figure}
    \centering
    \includegraphics[width=1\linewidth]{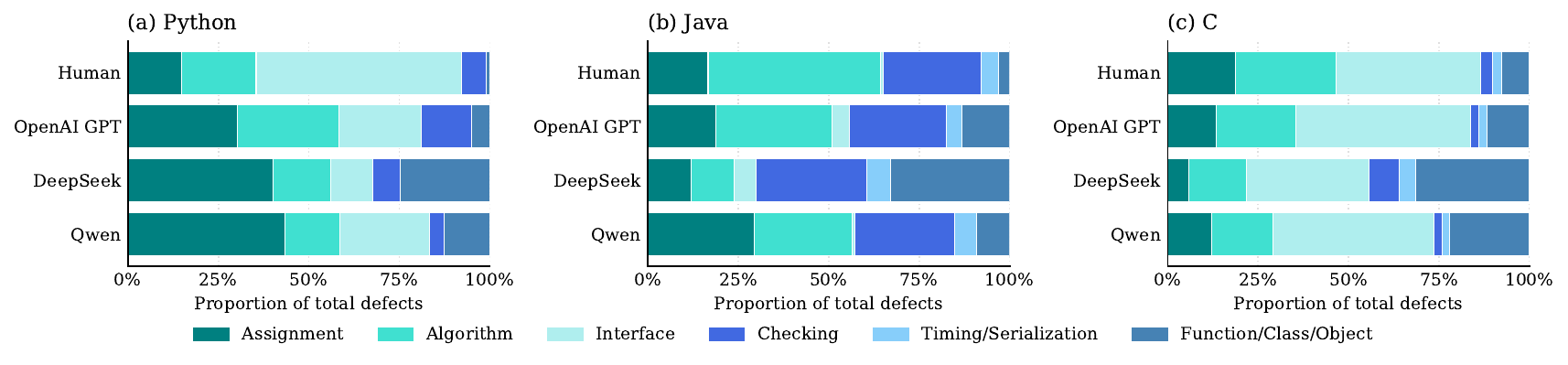}
    \caption{Distribution of ODC defect types across the four authors for each language.}
    \label{fig:odc_distribution}
\end{figure}

\tableautorefname~\ref{tab:defects_summary} summarizes the distribution of code issues in terms of \emph{defective samples}, \ie samples containing at least one detected issue, \emph{incorrect Samples}, \ie samples that are either syntactically invalid or entirely missing (\eg empty response due to incorrect interpretation of the input by the model), and \emph{total defects}, \ie cumulative issue count, since a single sample may contain multiple defects. \figureautorefname~\ref{fig:odc_distribution} shows the proportional distribution of ODC defect types across the four authors for each language. Tables~\ref{tab:top_defects_python}--\ref{tab:top_defects_c} report the five most-frequent specific defects per author, useful for interpreting which defect classes within each ODC category drive the patterns visible in \figureautorefname~\ref{fig:odc_distribution}.

\begin{table}[t]
\centering
\caption{Defect statistics by language and code author. Blue are \textcolor{blue}{\textbf{best}} scores, red are the \textcolor{red}{\textbf{worst}}.}
\small
\label{tab:defects_summary}
\begin{tabular}{
>{\raggedright\arraybackslash}m{2cm}
>{\raggedright\arraybackslash}m{1.5cm}
>{\raggedleft\arraybackslash}m{2.5cm}
>{\raggedleft\arraybackslash}m{2.5cm}
>{\raggedleft\arraybackslash}m{2cm}
}
\toprule
\textbf{Language} & \textbf{Author} & \textbf{Defective Samples} & \textbf{Incorrect Samples} & \textbf{Total Defects} \\
\midrule
\multirow{4}{*}{\makecell{Python \\(285,249)}}
  & Human       & 158,221 & 3,712  &  429,247 \\
  & OpenAI GPT  & \textcolor{blue}{\textbf{110,528}} & \textcolor{blue}{\textbf{2,549}}  &  \textcolor{blue}{\textbf{185,110}} \\
  & DeepSeek    & 173,899 & \textcolor{red}{\textbf{20,783}} & 319,643 \\
  & Qwen        & \textcolor{red}{\textbf{178,225}} & 7,190  &  \textcolor{red}{\textbf{574,916}} \\
\midrule
\multirow{4}{*}{\makecell{Java \\(221,795)}}
  & Human       & 76,049 & \textcolor{blue}{\textbf{245}} & 202,129 \\
  & OpenAI GPT  & 78,241 & 2,056 & 161,056 \\
  & DeepSeek    & \textcolor{red}{\textbf{155,077}} & \textcolor{red}{\textbf{34,382}} & \textcolor{red}{\textbf{242,437}} \\
  & Qwen        & \textcolor{blue}{\textbf{60,389}} & 3,833 & \textcolor{blue}{\textbf{123,180}} \\
\midrule
\multirow{4}{*}{\makecell{C \\(280,518)}}
  & Human       & 128,748 & \textcolor{blue}{\textbf{11,585}} & 224,585 \\
  & OpenAI GPT  & \textcolor{blue}{\textbf{118,915}} & 12,866 & \textcolor{blue}{\textbf{190,828}} \\
  & DeepSeek    & \textcolor{red}{\textbf{138,321}} & 36,942 & \textcolor{red}{\textbf{277,955}} \\
  & Qwen        & 128,912 & \textcolor{red}{\textbf{45,295}} & 239,334 \\
\bottomrule
\end{tabular}
\end{table}

\paragraph{\textbf{Python.}}
In Python, the OpenAI GPT model produces the cleanest code overall: lowest defective sample count (110,528, 38.75\%), lowest incorrect sample count (2,549, 0.89\%), and lowest number of total defects (185,110). At the opposite end, Qwen produces the highest total defect count (574,916), driven by both the largest defective-sample rate (62.48\%) and a high rate of incorrect samples (2.52\%). DSC sits between these two, with a defective rate close to Qwen (60.96\%) but the highest share of incorrect sample among Python authors (7.29\%), reflecting more frequent generation failures: of its 20,783 incorrect samples, 19,612 are empty outputs, suggesting that DSC sometimes returns no code at all. Human-written Python code yields 429,247 total defects, more than ChatGPT and DSC but less than Qwen, with a defective share of 55.47\% and only 1.30\% incorrect samples.

The ODC distribution in \figureautorefname~\ref{fig:odc_distribution}(a) reveals a clear divergence between human and AI-generated Python. Human-written Python code is dominated by \textit{Interface} defects, accounting for over 56\% of all human findings. However, a significant portion of these defects ($\sim$47\%) stems from a single code issue detected by Pylint, \ie \texttt{protected-access}, which accounts for 202,259 violations. This issue refers to accessing class members prefixed with an underscore (\eg \texttt{\_internal\_method()}), which are considered protected by convention. In open-source repositories, especially those involving complex inheritance hierarchies or performance- sensitive designs, such practices are common and often intentional, even if discouraged. If we exclude these occurrences, the human total drops to roughly 227,000 defects, comparable to the OpenAI GPT total. 
The AI-driven authors all show very different profiles: OpenAI GPT, DSC, and Qwen are all dominated by \textit{Assignment}-related issues, with \texttt{unused-argument} alone contributing 32,992, 111,703, and 213,264 occurrences, respectively (Table~\ref{tab:top_defects_python}).
DSC and Qwen further accumulate \textit{Function/Class/Object} defects, primarily \texttt{too-few-public-methods} for DSC (64,279) and \texttt{redefined-outer-name} for Qwen (57,252), reflecting structural problems in class design and global-name reuse that human-written code rarely exhibits.

\begin{table*}[t]
\centering
\caption{Top-5 most-frequent specific defects per author for Python.}
\footnotesize
\label{tab:top_defects_python}
\begin{tabular}{lr@{\hspace{1.5em}}lr@{\hspace{1.5em}}lr@{\hspace{1.5em}}lr}
\toprule
\multicolumn{2}{c}{\textbf{Human}} & \multicolumn{2}{c}{\textbf{OpenAI GPT}} & \multicolumn{2}{c}{\textbf{DeepSeek}} & \multicolumn{2}{c}{\textbf{Qwen}} \\
\midrule
\texttt{protected-access}    & 202k & \texttt{unused-argument}        &  33k & \texttt{unused-argument}        & 112k & \texttt{unused-argument}      & 213k \\
\texttt{unused-argument}     &  27k & \texttt{no-else-return}         &  27k & \texttt{too-few-public-methods} &  64k & \texttt{protected-access}     & 106k \\
\texttt{no-else-return}      &  23k & \texttt{unused-variable}        &  14k & \texttt{no-else-return}         &  33k & \texttt{redefined-outer-name} &  57k \\
\texttt{unused-variable}     &  13k & \texttt{unspecified-encoding}   &  14k & \texttt{unspecified-encoding}   &  13k & \texttt{no-else-return}       &  35k \\
\texttt{raise-missing-from}  &  10k & \texttt{protected-access}       &   7k & \texttt{redefined-outer-name}   &  13k & \texttt{unnecessary-pass}     &  31k \\
\bottomrule
\end{tabular}
\end{table*}

The recurrence of \texttt{unused-argument} across all three AI models, in extremely large numbers, suggests that they frequently replicate parameter patterns from the docstring or function signature without actually using them in the generated body. Structural issues like \texttt{too-few-public-methods} and \texttt{redefined-outer-name} additionally indicate difficulties in generating coherent and contextually-integrated class hierarchies, which are not commonly observed in human-written code.

\paragraph{\textbf{Java.}}
In Java, DSC exhibits the worst performance from every point of view (\tableautorefname~\ref{tab:defects_summary}): highest total defects (242,437), most defective samples (69.92\%), and most incorrect outputs (15.50\%, of which 19,071 are empty). DSC produces problematic outputs in more than 85\% of generation attempts, suggesting it struggles significantly with Java's stricter syntax and structural constraints; we note, however, that its most frequent issues (\tableautorefname~\ref{tab:top_defects_java}) refer to best practices and design defects rather than critical errors. At the other end, Qwen produces the highest quality Java code overall, with the lowest defective sample rate (27.23\%) and the lowest total defect count (123,180). Human-written Java yields 202,129 defects across 34.29\% of defective samples, with only 245 incorrect cases. ChatGPT ranks between these extremes (35.28\% defective).

The ODC distribution in \figureautorefname~\ref{fig:odc_distribution}(b) highlights different failure modes. Human-written Java is dominated by Algorithm defects (96,418 of 202,129, $\sim$48\%), led by \texttt{CyclomaticComplexity} (15,435), \texttt{AvoidInstantiatingObjectsInLoops} (14,351), \texttt{CognitiveComplexity} (13,065). These reflect maintainability concerns characteristic of mature, real-world codebases: overly complex control flow, algorithmic flaws, and improper exception handling. The DSC profile diverges sharply: it accumulates 79,724 \textit{Function/Class/Object} defects (the largest of any author across any language in our corpus) and 74,252 Checking defects, with \texttt{ImmutableField} (41,669) and \texttt{SystemPrintln} (37,986) as its top two violations. This indicates that DSC frequently defines fields that could be marked final, overuses low-level debugging constructs like \texttt{System.out.println()} and \texttt{printStackTrace()}, and includes unused or unreferenced fields and parameters, suggesting syntactically plausible but structurally shallow code with limited architectural awareness. Qwen and ChatGPT show more even spreads across categories, with Qwen's most frequent violation being \texttt{UnusedFormalParameter} (24,660) and ChatGPT's being \texttt{SystemPrintln} (10,428). These point to common problems in low-complexity utility methods and boilerplate code, where unused inputs and basic best-practice violations occur frequently.

\begin{table*}[t]
\centering
\caption{Top-5 most-frequent specific defects per author for Java.}
\footnotesize
\label{tab:top_defects_java}
\begin{tabular}{lr@{\hspace{1em}}lr@{\hspace{1em}}lr@{\hspace{1em}}lr}
\toprule
\multicolumn{2}{c}{\textbf{Human}} & \multicolumn{2}{c}{\textbf{OpenAI GPT}} & \multicolumn{2}{c}{\textbf{DeepSeek}} & \multicolumn{2}{c}{\textbf{Qwen}} \\
\midrule
\texttt{CyclomaticComplexity}             & 15k & \texttt{SystemPrintln}                    & 10k & \texttt{ImmutableField}        & 42k & \texttt{UnusedFormalParameter}            & 25k \\
\texttt{GuardLogStatement}                & 14k & \texttt{AvoidInstantiatingObjsInLoops} &  9k & \texttt{SystemPrintln}         & 38k & \texttt{SystemPrintln}                    & 13k \\
\texttt{AvoidInstantiatingObjsInLoops} & 14k & \texttt{UseVarargs}                       &  9k & \texttt{UnusedPrivateField}    & 16k & \texttt{UseVarargs}                       &  6k \\
\texttt{CognitiveComplexity}              & 13k & \texttt{LiteralsFirstInComparisons}       &  8k & \texttt{AvoidPrintStackTrace}  & 14k & \texttt{AppendCharacterWithChar}          &  6k \\
\texttt{AvoidCatchingGenericException}    & 10k & \texttt{UnusedLocalVariable}              &  7k & \texttt{UnusedFormalParameter} &  9k & \texttt{AvoidInstantiatingObjsInLoops} &  6k \\
\bottomrule
\end{tabular}
\end{table*}

\paragraph{\textbf{C}}
The C results (\tableautorefname~\ref{tab:defects_summary}) broadly mirror the behavior observed in Python and Java: OpenAI's gpt-oss is again the author generating the least amount of defective samples (118,915, 42.39\%) and total defects (190,828), while DSC confirms the last position, with the most total defects (277,955), driven by its highest defective rate (138,321, 49.31\%); and human-written C code is second best with 128,748 defective functions and $\sim$1.7 issues per sample. 
As for incorrect samples, Qwen takes the worst share (45,295, 16.15\%), whereas DSC held that position in Python and Java; Qwen's elevated rate is driven roughly equally by empty outputs (18,541, 6.61\%) and parse failures (26,754, 9.54\%), the latter reflecting frequent generation of malformed or truncated functions and occasional language drift toward C++. Human-written C produces the lowest incorrect rate among the four C authors (11,585, 4.13\%).

The ODC distribution in \figureautorefname~\ref{fig:odc_distribution}(c) reveals a C-specific concentration of defects in the \textit{Interface} category, which accounts for between 34\% and 48\% of total defects per author. For human, gpt-oss, and Qwen code this is overwhelmingly driven by a single Clang-Tidy check, \texttt{bugprone-easily-swappable-parameters}, which fires when a function declares two adjacent parameters of the same type. This pattern is structurally embedded in C's standard library and common API conventions (\eg \texttt{memcpy(void *dst, void *src, size\_t n)} and \texttt{strncpy(char *dst, const char *src, size\_t n)}), and accounts for 96.2\% of human \textit{Interface} defects, 94.1\% of gpt-oss, and 88.2\% of Qwen. DSC is the exception: this issue accounts for only 55.5\% of its \textit{Interface} category, with the remainder dominated by \texttt{cppcoreguidelines-avoid-non-const-global-variables} (38,656 instances, far above any other author), which flags mutable variables declared at file or namespace scope. Such globals introduce hidden inter-function dependencies, complicate reasoning about side effects, and undermine testability.

\begin{table*}[t]
\centering
\caption{Top-5 most-frequent specific defects per author for C.}
\footnotesize
\label{tab:top_defects_c}
\begin{tabular}{lr@{\hspace{1em}}lr@{\hspace{1em}}lr@{\hspace{1em}}lr}
\toprule
\multicolumn{2}{c}{\textbf{Human}} & \multicolumn{2}{c}{\textbf{OpenAI GPT}} & \multicolumn{2}{c}{\textbf{DeepSeek}} & \multicolumn{2}{c}{\textbf{Qwen}} \\
\midrule
\texttt{easily-swappable-params} & 86k & \texttt{easily-swappable-params} & 86k & \texttt{unused-params}                & 88k & \texttt{easily-swappable-params} & 94k \\
\texttt{no-int-to-ptr}           & 27k & \texttt{unused-params}           & 23k & \texttt{easily-swappable-params}      & 52k & \texttt{unused-params}           & 53k \\
\texttt{signed-bitwise}          & 25k & \texttt{signed-bitwise}          & 21k & \texttt{avoid-non-const-global-vars}  & 39k & \texttt{no-int-to-ptr}           & 18k \\
\texttt{branch-clone}            & 22k & \texttt{no-int-to-ptr}           & 16k & \texttt{signed-bitwise}               & 20k & \texttt{signed-bitwise}          & 16k \\
\texttt{unused-params}           & 17k & \texttt{branch-clone}            & 14k & \texttt{err33-c}                      & 17k & \texttt{branch-clone}            & 13k \\
\bottomrule
\end{tabular}
\end{table*}

The Top-5 specific defects for C generated by OpenAI GPT and Qwen overlap exactly with the ones frequently encountered in human code, but the magnitudes diverge in a consistent direction across all three LLMs. The defect that grows most sharply is \texttt{misc-unused-parameters}: human C emits 17,435 instances (7.8\% of human Total Defects), OpenAI GPT emits 22,567 (11.8\%), Qwen emits 52,932 (22.1\%), and DSC emits 87,606 (31.5\%). The same monotonic pattern, with DSC at the extreme, shows up in unused-arguments in Python (\tableautorefname~\ref{tab:top_defects_python}) and unused formal parameters in Java (\tableautorefname~\ref{tab:top_defects_java}), suggesting a shared LLM failure mode in which generated functions declare parameters that the function body does not subsequently use, presumably because the model replicates parameters from the prompt's signature or docstring without integrating them into the generated body.

DSC's profile is once again the outlier even within this shared LLM pattern. \texttt{cppcoreguidelines-avoid-non-const\--global-variables} (38,656 instances) appears as DSC's third most-frequent issue, far above any other author, indicating a recurring tendency in DSC to expose state across the module boundary rather than keeping it local to the generating function. DSC also shows substantially higher \textit{Checking}-category defects (22,698 vs.\ 4,360--7,583 for the other authors), driven by missing return-value checks (\texttt{cert-err33-c}, 17,010 instances), suggesting that DSC produces C code that calls functions whose return values it does not verify.

OpenAI GPT and Qwen show milder forms of the same generation behavior. OpenAI's Top-5 distribution in C is the closest to human in magnitude. Qwen sits between OpenAI and DSC: its \texttt{misc-unused-parameters} count is 3$\times$ the human baseline, but it does not exhibit the global-variable or unchecked-return-value patterns that distinguish DSC. Together, these patterns indicate a graded LLM failure mode: all three models over-declare unused parameters, DSC most severely. %, and DSC additionally over-decorates generated functions with auxiliary globals and unchecked external calls that neither OpenAI GPT nor Qwen produce at comparable rates. This generalizes the ``auxiliary structure'' pattern observed for DSC's Python and Java output (where the auxiliary structures took the form of unused fields, unused private fields, and structurally redundant class members) to a C-specific instantiation in which the auxiliary structures are unused parameters, global variables, and unchecked function returns.

\begin{mainbox}{}
Human-written and AI-generated code differ structurally, not just quantitatively. Across Python, Java, and C, the human most frequent defects are dominated by classes characteristic of mature codebases (control-flow complexity in Java, exception handling and protected-member access in Python, buffer-related issues in C), reflecting the trade-offs of evolved real-world software. AI-generated code, by contrast, concentrates on simpler, repetitive patterns: unused parameters, structural redundancy, and surface-level formatting violations. This profile difference is consistent across all four LLMs but graded in severity, with DSC the most extreme and OpenAI GPT the closest to human; Qwen sits between them, with most of its surplus concentrated at the parsing boundary (incorrect samples) rather than within generated code. Across languages, the absolute volume of defects is comparable between human and AI authors in C, but substantially higher in Python and Java, indicating that the more permissive the language ecosystem, the more amplified the AI-vs-human gap.
\end{mainbox}

\subsection{RQ$_3$: Do security vulnerabilities differ between human-written and AI-generated code across programming languages, in terms of type and severity?}
\label{sec:results-vulns}

\begin{table}[t]
\centering
\caption{Security vulnerabilities statistics by language and code author. Blue are \textcolor{blue}{\textbf{best}} scores, red are the \textcolor{red}{\textbf{worst}}.}
\small
\label{tab:vulnerabilities_summary}
\begin{tabular}{
>{\raggedright\arraybackslash}m{2cm}
>{\raggedright\arraybackslash}m{1.5cm}
>{\raggedleft\arraybackslash}m{2.5cm}
>{\raggedleft\arraybackslash}m{2.5cm}
>{\raggedleft\arraybackslash}m{2cm}
}
\toprule
\textbf{Language} & \textbf{Author} & \textbf{Vulnerable Samples} & \textbf{Unique CWEs} & \textbf{Total CWEs} \\
\midrule
\multirow{4}{*}{\makecell{Python \\(285,249)}}
  & Human       & \textcolor{blue}{\textbf{15,835}} & \textcolor{red}{\textbf{45}} & \textcolor{blue}{\textbf{25,678}} \\
  & OpenAI GPT  & 22,864 & \textcolor{red}{\textbf{45}} & \textcolor{red}{\textbf{40,035}} \\
  & DeepSeek    & \textcolor{red}{\textbf{22,878}} & \textcolor{blue}{\textbf{43}} & 39,982 \\
  & Qwen        & 16,316 & \textcolor{red}{\textbf{45}} & 34,250 \\
\midrule
\multirow{4}{*}{\makecell{Java \\(221,795)}}
  & Human       & \textcolor{blue}{\textbf{6,660}} & 42 & \textcolor{blue}{\textbf{11,677}} \\
  & OpenAI GPT  & 18,144 & 49 & 29,807 \\
  & DeepSeek    & \textcolor{red}{\textbf{43,386}} & \textcolor{red}{\textbf{51}} & \textcolor{red}{\textbf{76,678}} \\
  & Qwen        & 13,620 & \textcolor{blue}{\textbf{41}} & 21,301 \\
\midrule
\multirow{4}{*}{\makecell{C \\(280,518)}}
  & Human       & 24,976 & 28 & \textcolor{red}{\textbf{58,061}} \\
  & OpenAI GPT  & \textcolor{red}{\textbf{26,315}} & \textcolor{red}{\textbf{29}} & 48,222 \\
  & DeepSeek    & 25,551 & \textcolor{red}{\textbf{29}} & 49,907 \\
  & Qwen        & \textcolor{blue}{\textbf{18,839}} & \textcolor{blue}{\textbf{27}} & \textcolor{blue}{\textbf{36,973}} \\
\bottomrule
\end{tabular}
\end{table}

\begin{figure}
    \centering
    \includegraphics[width=1\linewidth]{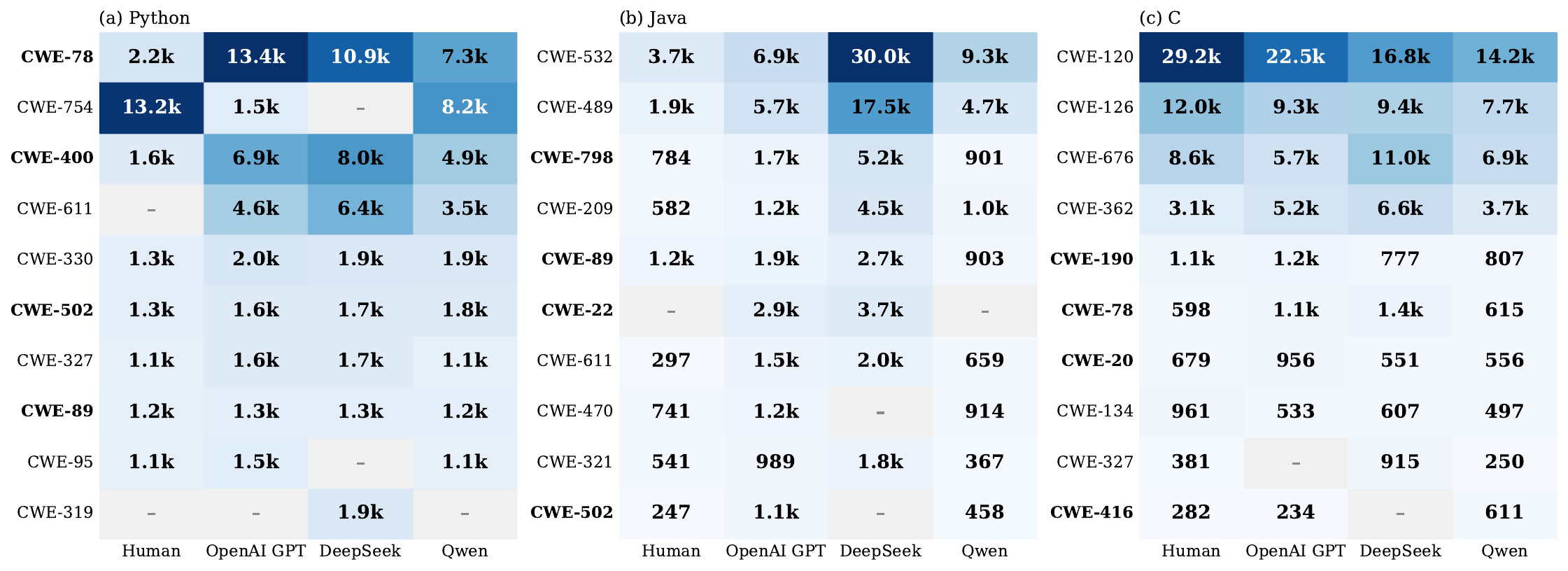}
    \caption{Heatmap of the top-10 CWEs distribution across code authors, per language. Each panel shows the union of the four authors' per-author top-10 CWEs, trimmed to the ten entries with the highest cross-author totals. Cells outside an author's top-10 are marked with a dash. CWEs belonging to MITRE's Top 25 are shown in bold.}
    \label{fig:heatmap_CWEs}
\end{figure}

Although AI code assistants may reliably generate syntactically correct and functionally valid code, recent evidence suggests that they may also propagate insecure coding patterns, often in subtle and systematic ways~\cite{pearce2025asleep}. Understanding whether the nature of these vulnerabilities diverges from those traditionally introduced by human developers is essential to adapting software security tools and mitigation frameworks accordingly.

To investigate differences in the nature and frequency of security vulnerabilities across human-written and AI-generated code, we analyze every code sample with Semgrep, configured with its curated security-focused rule sets, and map each finding to its associated CWE entry through Semgrep's rule metadata. We compare the four code authors per language: Human, OpenAI GPT (ChatGPT for Python and Java; gpt-oss for C), DSC, and Qwen. To increase interpretability of the results, this section provides an overview of our findings; complete per-author and per-CWE distributions are available in the replication package~\cite{replication}.

\tableautorefname~\ref{tab:vulnerabilities_summary} summarizes the security profile of each author in terms of \textit{vulnerable samples}, \ie samples containing at least one security finding, \textit{unique CWEs}, \ie the number of distinct CWE identifiers triggered, and \textit{total issues}, \ie the cumulative finding count across the corpus, since a single function can trigger multiple CWEs. \figureautorefname~\ref{fig:heatmap_CWEs} complements the analysis with a per-language breakdown of the most prominent CWEs. 
Rows are constructed by ranking, for each author, the CWEs by total vulnerabilities count; the union of the four per-author top-10 lists is then trimmed to ten entries by cross-author total. Cells where the CWE is not in that author's top-10 are rendered as a dash (``--''); the underlying count may still be non-zero but falls below the author's top-10 cutoff. %, so a dash should be read as ``outside this author's top-10'', not as ``zero occurrences''. 
CWEs belonging to MITRE's Top 25 most dangerous software weaknesses~\cite{top25mitre} are bold.

\paragraph{\textbf{Python.}} In Python, DSC and OpenAI GPT produce comparable volumes of vulnerable code (22,878 and 22,864 vulnerable samples, both at $\sim$8.02\% of the dataset), with OpenAI GPT marginally higher on total issues (40,035 vs 39,982) and CWE breadth (45 vs. 43 unique types). The two LLMs are nearly indistinguishable at the corpus level, and both substantially exceed the human baseline of 15,835 vulnerable samples (5.55\%) and 25,678 total issues. Qwen is the closest LLM to the human baseline on count (16,316 vulnerable samples, 5.72\%; 34,250 total issues) but exhibits the highest issue density of the four authors: 2.10 issues per vulnerable function, against 1.75 for OpenAI GPT and DSC and 1.62 for human.

\figureautorefname~\ref{fig:heatmap_CWEs}(a) shows that the LLM surplus is concentrated in a small set of CWE classes. CWE-78 (OS Command Injection) reaches 13,419 instances in OpenAI GPT and 10,908 in DSC, against 2,243 in human code. CWE-400 (Uncontrolled Resource Consumption) reaches 7,987 in DSC and 6,887 in OpenAI GPT, against 1,631 in human. CWE-611 (XML External Entity Reference) is almost absent in human (267 instances, below the human top-10 cutoff) but reaches 6,372 in DSC and 4,587 in OpenAI GPT. The reverse skew is also present: CWE-754 (Improper Check for Unusual or Exceptional Conditions) dominates the human profile with 13,157 instances ($\sim$51\% of human total) but accounts for only 1,539 issues in OpenAI GPT and 528 in DSC, although Qwen reproduces it at 8,214. DSC alone among the four authors places CWE-319 (Cleartext Transmission of Sensitive Information) in its top-10 (1,928 instances), absent from the top-10 of every other Python author.

\paragraph{\textbf{Java.}} In Java, the security gap between human and AI-generated code is the widest in our study. DSC is the worst performer on every metric: 43,386 vulnerable samples (19.56\%), 76,678 total issues, and 51 distinct CWE types triggered, the highest single-author count across the three languages. ChatGPT follows with 18,144 vulnerable samples (8.18\%), 29,807 issues, and 49 unique CWEs. Qwen produces the cleanest LLM output, with 13,620 vulnerable samples (6.14\%) and the lowest issue density of the four Java authors (1.56 issues per vulnerable function). All three LLMs exceed the human baseline of 6,660 vulnerable samples (3.00\%).

The CWE distributions in \figureautorefname~\ref{fig:heatmap_CWEs}(b) reveal a shared LLM failure mode amplified to extreme levels in DSC. CWE-532 (Insertion of Sensitive Information into Log File) is the top CWE for every Java author, but the magnitudes diverge sharply: human at 3,736, ChatGPT at 6,915, Qwen at 9,271, and DSC at 30,031, an 8$\times$ surplus over human. CWE-489 (Active Debug Code) follows the same pattern (human 1,948; DSC 17,533, a 9$\times$ surplus). These two CWEs jointly account for 47,564 of DSC's 76,678 issues (62.0\%), indicating that DSC's Java output frequently mixes production code with logging, debugging, and stack-trace patterns appropriate only to development builds. CWE-798 (Use of Hard-coded Credentials, 5,199 in DSC vs 784 in human), CWE-209 (Sensitive Information in Error Messages, 4,549 vs 582), and CWE-22 (Path Traversal, 3,738 in DSC and 2,886 in ChatGPT against 14 in human and 4 in Qwen, both below their respective top-10 cutoffs) are similarly amplified. Qwen tracks the human shape more closely than DSC or ChatGPT: it inflates CWE-532, CWE-489, CWE-209, and CWE-798 by smaller multiples than the other LLMs, and stays at or below the human baseline on CWE-89 and CWE-22.

\paragraph{\textbf{C}} The pattern observed in Python and Java does not extend to C. Human-written C is no longer the safest code by total issue count: it produces 58,061 issues across 24,976 vulnerable samples (8.90\%), with an issue density of 2.32 issues per vulnerable function, higher than any LLM in our study. Among the LLMs, OpenAI GPT yields the most vulnerable samples (26,315, 9.38\%), DSC sits at 25,551 (9.11\%), and Qwen has the lowest vulnerable count (18,839, 6.72\%) and the lowest total issue count (36,973). However, Qwen's apparent advantage is partly an artifact of its 16.15\% incorrect-sample rate for C (\tableautorefname~\ref{tab:defects_summary}), the highest among C authors and roughly $3.5\times$ the rate for human (4.13\%) or gpt-oss (4.59\%); Semgrep cannot raise findings on functions it cannot parse. The unique-CWE counts are tightly clustered across authors (27 to 29), reflecting that C has a smaller but more concentrated vulnerability surface than Python or Java: a handful of memory-safety patterns drive the bulk of findings.

\begin{figure}[t]
\centering
\includegraphics[width=0.8\linewidth]{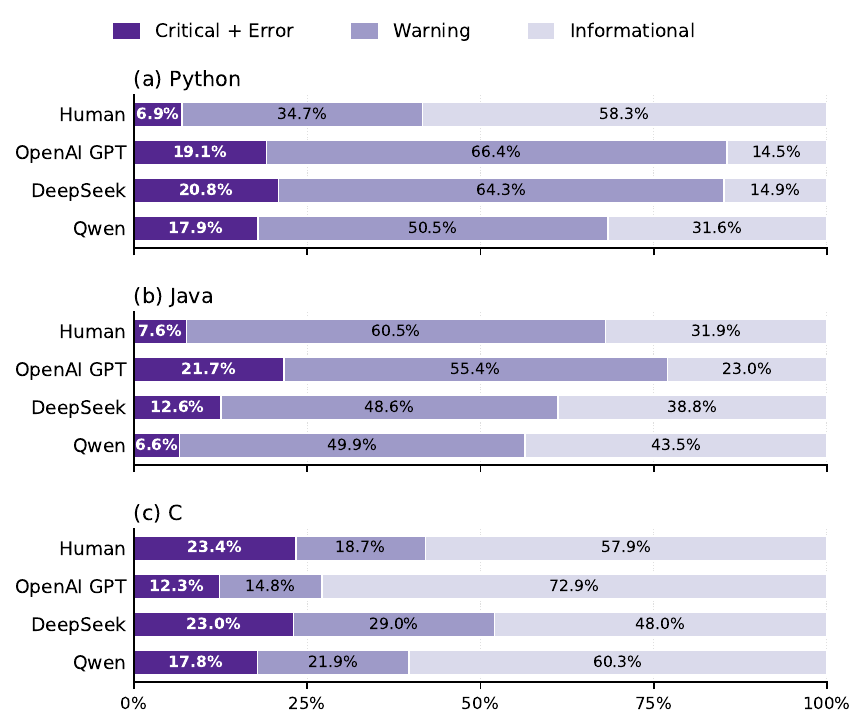}
\caption{Severity composition of Semgrep findings across code authors for (a)~Python, (b)~Java, (c)~C. Each bar is normalized to 100\% of that author's total Semgrep findings. The CRITICAL and ERROR severities are merged into a single ``Critical~+~Error'' bucket because CRITICAL findings are rare in our corpus (no more than 26 per author/language combination).}
\label{fig:severity-profile}
\end{figure}

\figureautorefname~\ref{fig:heatmap_CWEs}(c) makes the composition explicit. Human-written C is dominated by classic memory-safety bugs: CWE-120 (Classic Buffer Overflow) accounts for 29,164 of the 58,061 human issues (50.2\%), followed by CWE-126 (Buffer Over-read, 12,035) and CWE-676 (Use of Potentially Dangerous Function, 8,606). These three CWEs jointly account for 85.8\% of human C findings. All three LLMs reduce the dominant CWE-120 count substantially: gpt-oss drops it to 22,454 ($-$23\% vs human), DSC to 16,786 ($-$42\%), Qwen to 14,231 ($-$51\%). CWE-126 and CWE-676 follow the same reduction trend for gpt-oss and Qwen, with DSC the exception on CWE-676: at 11,027 instances it exceeds even the human baseline.

The memory-safety reduction does not translate into safer C code overall. Two compensating effects are visible. First, race conditions (CWE-362, Time-of-check Time-of-use) are more frequent in every LLM than in human code: gpt-oss at 5,231 and DSC at 6,593, against 3,089 for human; Qwen is the only LLM close to the human level at 3,739. Second, OS command injection (CWE-78) is amplified by DSC (1,450 instances) and gpt-oss (1,088) over the human baseline of 598. The result is a redistribution rather than an elimination of risk: LLMs trade fewer buffer-related bugs for more concurrency and injection flaws, with DSC again the most extreme on both shifts.

\paragraph{\textbf{Severity profile.}} 
The CWE-type comparison so far weights all Semgrep findings equally, but Semgrep's findings also come with different severity levels which reflect how directly exploitable a pattern is: \texttt{Critical} and \texttt{Error} flag patterns whose risk is clear regardless of surrounding context, while \texttt{Warning} and \texttt{Informational} flag context-dependent issues that may or may not yield a working exploit in practice. The same CWE label can therefore aggregate findings whose practical risk differs substantially. To disentangle high-severity volume from lower-severity volume, we examine how each author's vulnerable code distribute across severity buckets. Normalized results are shown in \figureautorefname~\ref{fig:severity-profile}.

Three patterns are not visible in the raw CWE counts. First, Python LLMs concentrate findings in the \texttt{Warning} bucket (50.5\% to 66.4\% of findings, against 34.7\% for human Python), while human Python is heavily skewed toward \texttt{Informational} (58.3\%, driven almost entirely by CWE-754). Second, Java Qwen's severity profile is the closest of any LLM to human across the three languages: its high-severity share (6.6\%) is the only LLM value below the corresponding human baseline (7.6\%), with a high-severity count of 1,409 against the human 889 (a $1.6\times$ ratio, against the $7.3\times$ for ChatGPT and $10.9\times$ for DSC). The Java security gap between Qwen and human concentrates in the \texttt{Warning} and \texttt{Informational} buckets rather than in critical issues. Third, C is the only language where the human high-severity count exceeds every LLM: human at 13,565 \texttt{Critical+Error} findings against gpt-oss 5,941 ($-$56\%), Qwen 6,580 ($-$51\%), and DSC 11,491 ($-$15\%). The memory-safety reduction reported above (CWE-120, CWE-126) is therefore concentrated in the \texttt{Error}-rated patterns that the static analyzers flag as most exploitable issues.

\begin{mainbox}{}
The security comparison between human and AI-generated code is language-dependent on two axes simultaneously: volume and severity. In Python and Java, LLMs produce $1.4\times$ to $6.5\times$ more vulnerable samples than human, with high-severity (Critical+Error) findings inflated by $1.6\times$ (Qwen Java) to $10.9\times$ (DSC Java); the surplus skews toward injection flaws (CWE-78 in Python, CWE-22 in Java), insecure configurations (CWE-489 in Java), and information exposure (CWE-532 in Java, CWE-319 unique to DSC Python). In C, the ordering reverses for high-severity findings: human produces 13,565 Critical+Error findings against gpt-oss 5,941 ($-$56\%) and Qwen 6,580 ($-$51\%), almost entirely driven by Error-rated buffer-overflow patterns (CWE-120); LLMs redistribute the residual risk toward race conditions (CWE-362) and OS command injection (CWE-78). DSC is the worst performer in Python and Java but sits mid-pack in C; Qwen is the closest LLM to the human baseline in Python and Java, and the only LLM whose Java high-severity count is within twice the human one. Within each language the same handful of CWEs dominates every author's top-10, but magnitudes vary by an order of magnitude. The CWE-class and severity dimensions support the same conclusion: AI-generated code requires language-specific security assessment rather than a single cross-language risk model.
\end{mainbox}

\subsection{RQ4: Do the structural and statistical properties of code (RQ1) explain differences in defect and vulnerability profiles (RQ2, RQ3) across programming languages?}
\label{sec:results-correlations}

\begin{figure*}[t]
  \centering
  \includegraphics[width=\textwidth]{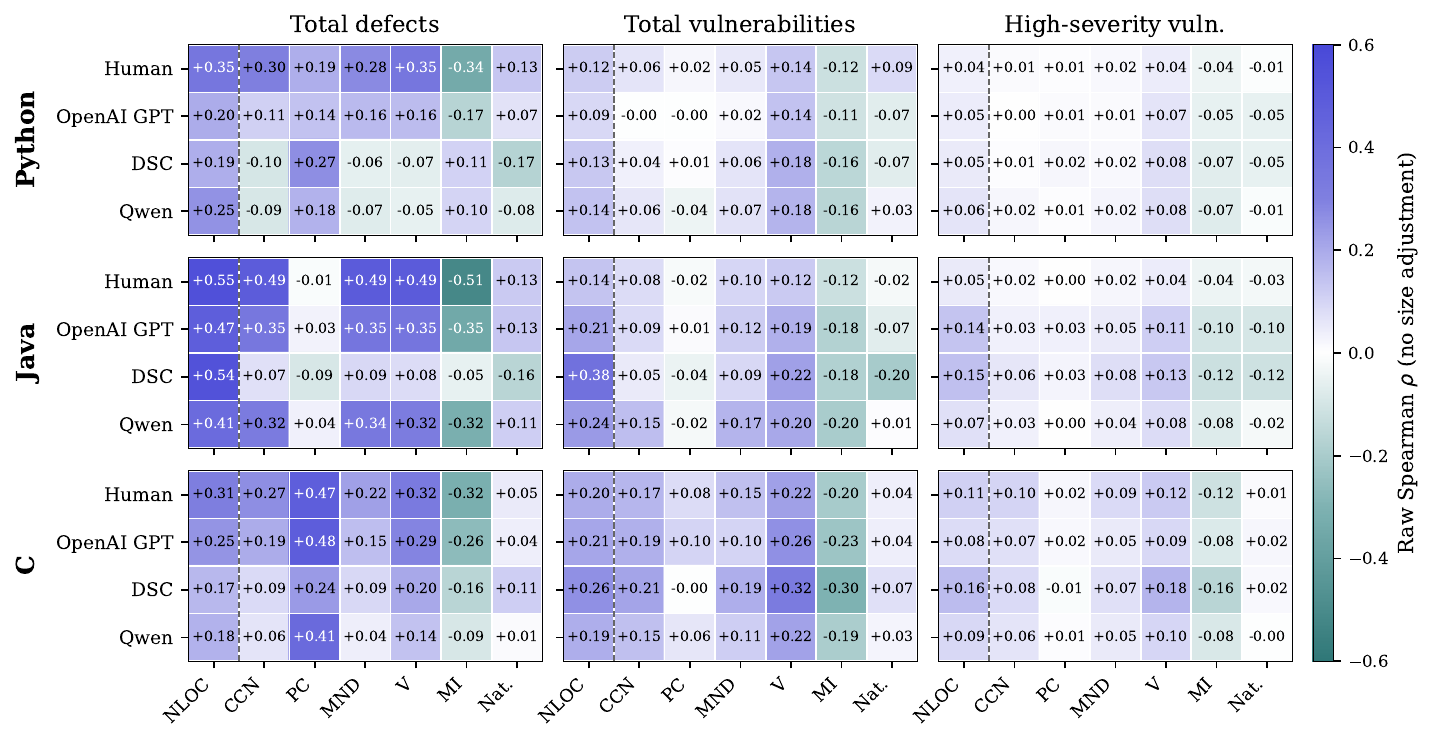}
  \caption{Raw Spearman correlations between structural and statistical predictors and
  aggregate outcomes (total defects, total vulnerabilities, high-severity
  vulnerabilities), per language and code author, without size adjustment. The NLOC
  column is the whole-sample NLOC, i.e., the variable used as the adjustment
  control in \figureautorefname~\ref{fig:rq4_adjusted}, separated by the dashed line.}
  \label{fig:rq4_raw}
\end{figure*}

\begin{figure*}[t]
  \centering
  \includegraphics[width=\textwidth]{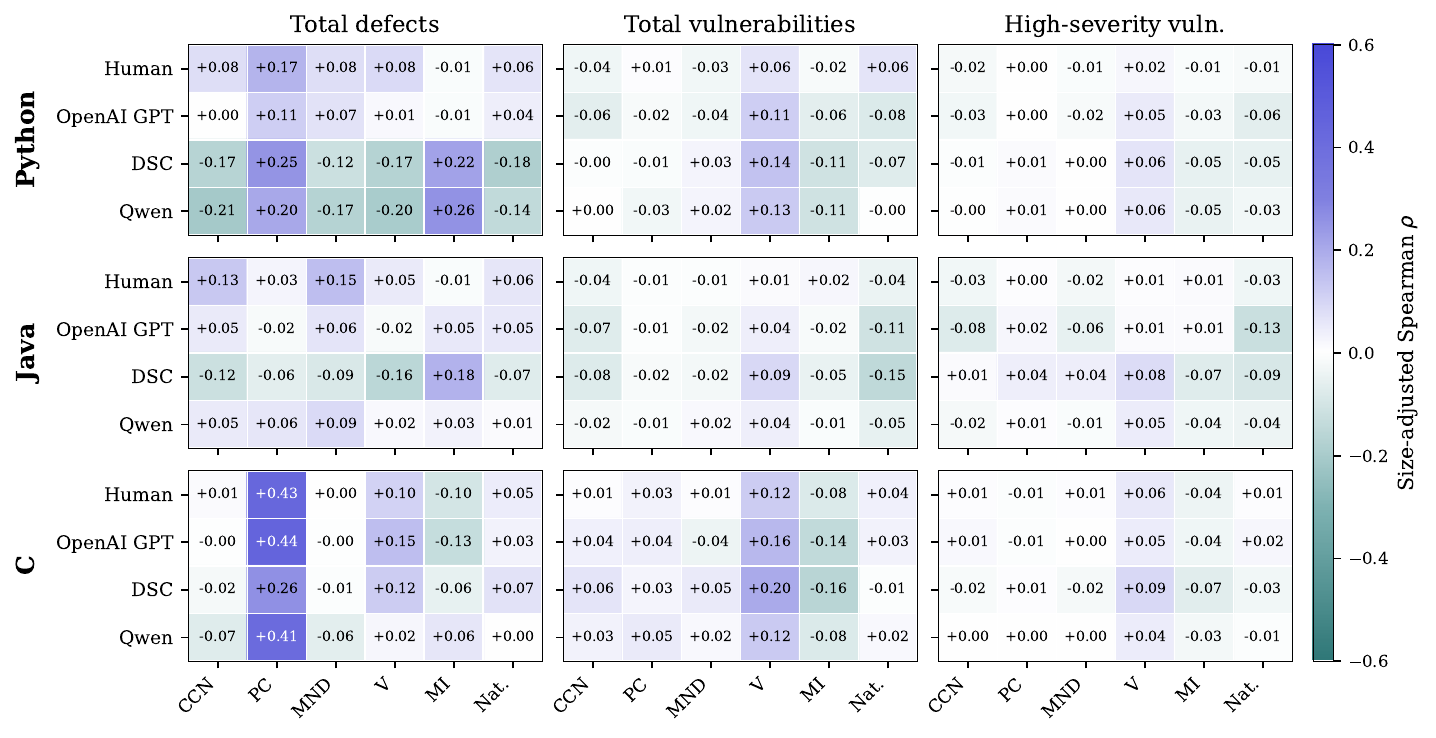}
  \caption{Size-adjusted Spearman correlations (partialled on whole-sample NLOC)
  between predictors and aggregate outcomes, per language and code author. Whole-sample
  NLOC is the adjustment variable and does not appear as a predictor.}
  \label{fig:rq4_adjusted}
\end{figure*}

\begin{figure*}[t]
  \centering
  \includegraphics[width=\textwidth]{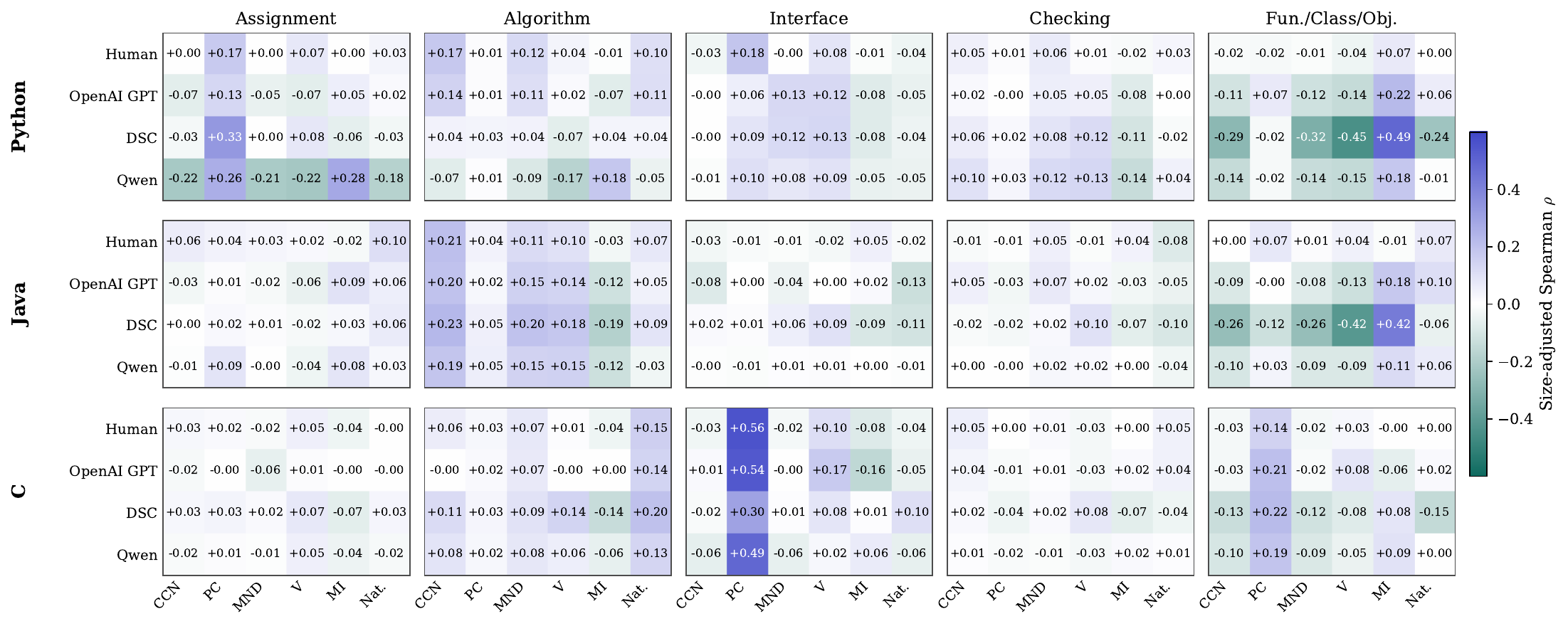}
    \caption{Size-adjusted Spearman correlations disaggregated by ODC defect type.}
  \label{fig:rq4_odc_grid}
\end{figure*}

\begin{figure*}[t]
  \centering
  \includegraphics[width=\textwidth]{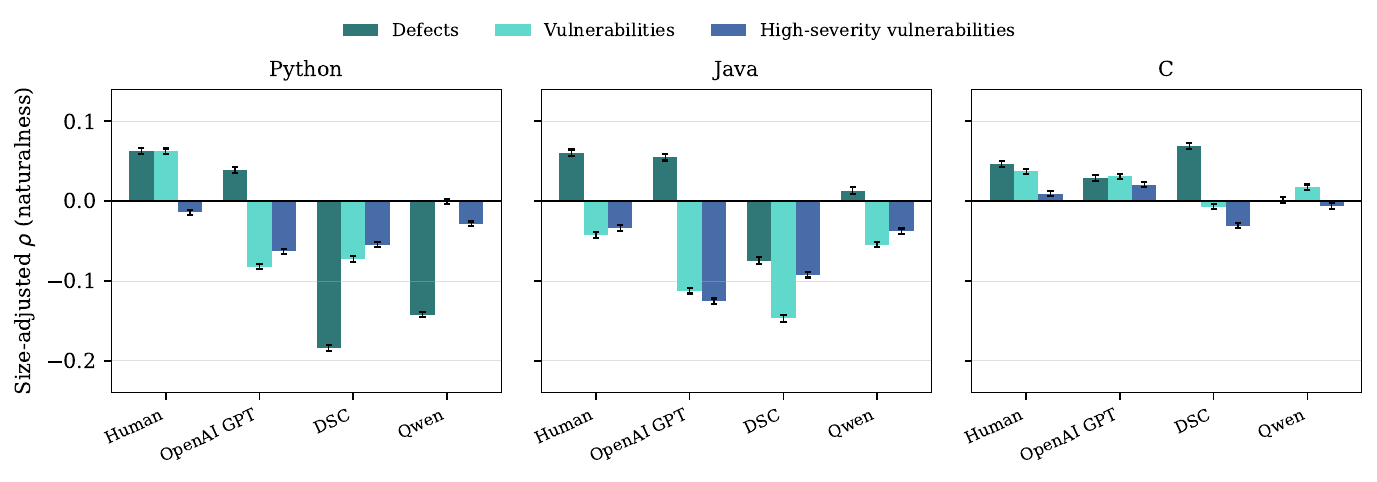}
  \caption{Size-adjusted correlation between structural naturalness (identifier-normalized cross-entropy; higher = less predictable) and defect, vulnerability, and high-severity vulnerability counts, per language and author, with 95\% bootstrap confidence intervals.}
  \label{fig:rq4_naturalness}
\end{figure*}

\begin{figure*}[t]
  \centering
  \includegraphics[width=\textwidth]{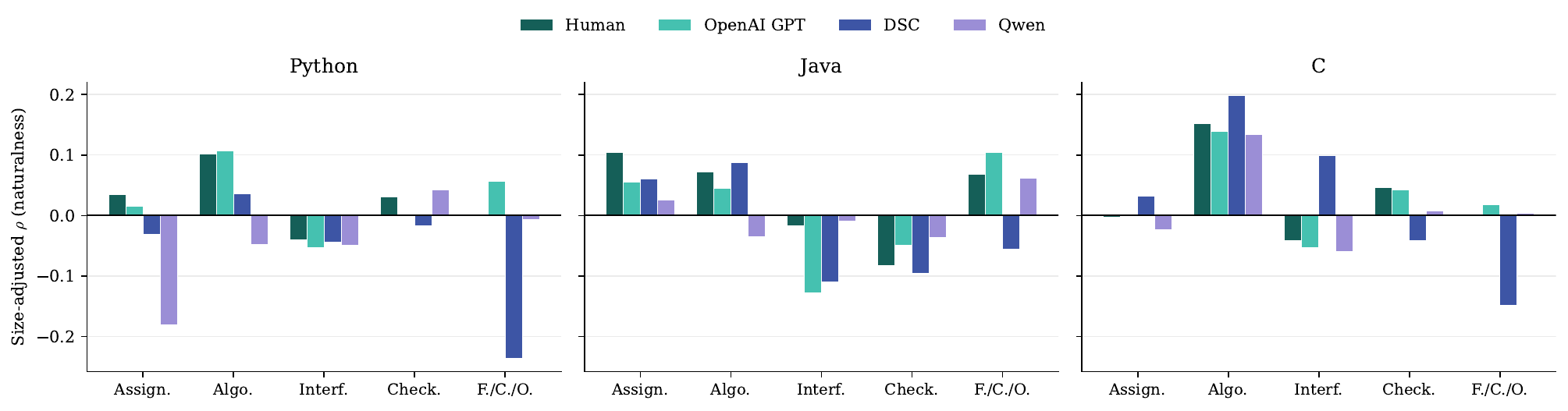}
  \caption{Size-adjusted naturalness--defect correlation by ODC type. DSC's Python
  reversal concentrates in Function/Class/Object, Qwen's in Assignment; in C, Algorithm
  is the single category on which all four authors couple positively.}
  \label{fig:rq4_naturalness_odc}
\end{figure*}

To answer RQ$_4$, we test whether the structural and statistical properties analyzed in RQ1 are associated with the defect and vulnerability profiles observed in RQ$_2$ and RQ$_3$. For each language and code author, we compute Spearman correlations between predictors and outcomes. The predictors include representative metrics for size and structure (NLOC, CCN, Parameter Count, Maximum Nesting Depth, Halstead Volume, Maintainability Index) and statistical naturalness. The outcomes include aggregate counts of defects, vulnerabilities, and high-severity vulnerabilities, as well as ODC-disaggregated defect counts. We report each relationship in two forms: raw correlations, and size-adjusted correlations that control for whole-sample NLOC. The adjustment separates associations driven by code volume from associations that remain after accounting for size. We report 95\% bootstrap confidence intervals over 1,000 resamples and interpret correlations by magnitude, direction, and stability across languages and authors. All results are associational and should not be interpreted causally.

Two measurement details matter for interpretation: \textit{(i)} the size control is the whole-sample NLOC; \textit{(ii)} the naturalness predictor is the identifier-normalized (structural) cross-entropy of \S\ref{sec:naturalness-design}, scored under each author's own 6-gram model on held-out folds, in bits per token; higher values indicate less locally predictable structure.

The raw correlations in \figureautorefname~\ref{fig:rq4_raw} show that NLOC is positively associated with total defects for every author in every language. The effect is strongest in Java, where the raw NLOC--defect correlation reaches $+0.55$ for human code and $+0.54$ for DSC, but it is visible across Python and C as well. Other size-related metrics show similar behavior before adjustment: CCN, MND, and Halstead Volume are often positively correlated with total defects, while Maintainability Index is negatively correlated because it is constructed to decrease as size and complexity increase.
After controlling for NLOC, most of this signal disappears. In \figureautorefname~\ref{fig:rq4_adjusted}, the adjusted correlations between structural predictors and total defects are generally small, especially for human and OpenAI GPT code. For example, Java human code moves from raw correlations around +0.49 for CCN, MND, and
Halstead Volume to adjusted correlations of $+0.13$, $+0.15$, and $+0.05$; the corresponding MI association moves from $-0.51$ to $-0.01$. A similar collapse occurs for Java OpenAI GPT and for human code in Python and C. Security outcomes are even less explained by structural metrics: after size adjustment, no aggregate predictor exceeds $|\rho|=0.21$ for total or high-severity vulnerabilities across any author and language. 
\emph{These results indicate that conventional complexity metrics are not reliable proxies for the defect and vulnerability profiles. Much of their raw association reflects exposure: larger samples contain more code and therefore more opportunities for authors to inject issues.}

The ODC-disaggregated view in \figureautorefname~\ref{fig:rq4_odc_grid} shows that only a few associations survive size adjustment with moderate magnitude, and most of them are localized to specific authors, languages, and defect types rather than recurring across the full grid. The first is a DSC-specific Function/Class/Object cluster in Python and Java. Code generated by DeepSeek-Coder that has lower internal structural complexity exhibits more Function/Class/Object defects. In Python, the adjusted correlations are $-0.29$ for CCN, $-0.32$ for MND, and $-0.45$ for Halstead Volume; in Java, they are $-0.26$, $-0.26$, and $-0.42$, respectively. MI has the opposite sign ($+0.49$ in Python and $+0.42$ in Java), as expected from its inverse relationship with size and complexity. Naturalness points in the same direction in Python ($-0.24$), indicating that the most structurally predictable DSC outputs are also those most associated with Function/Class/Object defects. This pattern should not be read as a general rule that simpler code is more defective. Rather, it points to a specific generation habit of DSC: producing low-complexity class-like scaffolding or unused structural elements that trigger Function/Class/Object warnings. This interpretation is consistent with the RQ2 findings, where DSC's design-category warnings are dominated by class-scaffolding and unused-structure patterns. It is also supported by the absence of the same pattern in C, where the class-wrapper failure mode cannot occur: DSC's C Function/Class/Object correlations do not form the same negative complexity cluster.

A second residual signal appears in C Interface defects and is shared across all authors. Parameter Count is moderately to strongly associated with Interface findings, with adjusted correlations of $+0.56$ for human code, $+0.54$ for OpenAI GPT, $+0.30$ for DSC, and $+0.49$ for Qwen. Although this is the strongest cross-author structural association in C, its interpretation requires caution because the predictor and one dominant Interface rule are not independent. In particular, many C Interface findings are produced by the Clang-Tidy check \texttt{bugprone-easily-swappable-parameters}, which becomes applicable only when a function has multiple parameters of compatible types. The observed PC--Interface association therefore likely reflects, at least in part, the mechanics and coverage of this check: functions with more parameters provide more opportunities for adjacent same-type parameters and, consequently, for this Interface warning to fire. For this reason, we treat the association as rule-conditioned evidence rather than as a general claim that larger function interfaces are intrinsically lower quality. The weaker coefficient for DSC is also consistent with the RQ2 finding that DSC's Interface findings are less dominated by this check than those of the other authors.

The same grid contains smaller localized associations in Python Assignment defects, most notably the Parameter Count association for DSC ($+0.33$) and Qwen ($+0.26$). We interpret these cells with similar caution. They are confined to one language and one defect type, and they likely reflect the interaction between function signatures and rule-specific findings rather than a general structural explanation of defects. In particular, RQ2 shows that Assignment findings in generated Python code often involve unused arguments or signature-related boilerplate; functions with more parameters simply create more opportunities for such findings to occur.

The naturalness results in \figureautorefname~\ref{fig:rq4_naturalness} show a different pattern. After size adjustment, identifier-normalized cross-entropy is weakly but consistently positive for human total defects in all three languages: $+0.06$ in Python, $+0.06$ in Java, and $+0.05$ in C. In other words, less predictable human code tends to carry slightly more defects, a direction consistent with prior work linking naturalness and bug-proneness in human code~\cite{ray2016naturalness}. OpenAI GPT follows the same direction for total defects in all three languages, although with similarly small magnitudes. DSC and Qwen diverge from this pattern. DSC shows negative naturalness--defect associations in Python ($-0.18$) and Java ($-0.07$), while Qwen shows a negative association in Python ($-0.14$) and a near-zero association in Java. In C, the divergence largely disappears: all authors are near zero or weakly positive for total defects. This suggests that the relationship between predictability and defects is not simply a human-versus-AI split. Instead, it is model- and language-dependent. \emph{For human code, lower predictability tends to align with higher defect density. For generated code, the opposite can occur because the most predictable outputs are not necessarily the cleanest ones; they may be templated patterns that carry recurring defects.}

The ODC-level naturalness results in \figureautorefname~\ref{fig:rq4_naturalness_odc} explain where this reversal comes from. DSC's negative Python association is concentrated in Function/Class/Object defects, where the naturalness correlation reaches approximately $-0.24$. This matches the low-complexity scaffolding mechanism identified above: the most structurally predictable DSC outputs are also those that trigger design-category warnings. Qwen's strongest negative Python association appears in Assignment, around $-0.18$, consistent with boilerplate code and unused-argument findings. In C, by contrast, Algorithm is the only ODC category for which all authors show a positive association between naturalness and defects, ranging from weak to moderate values. This suggests that, when the boilerplate mechanisms observed in Python and Java are less dominant, the human-like direction re-emerges. Naturalness carries little aggregate signal for security outcomes. Human correlations with total and high-severity vulnerabilities remain close to zero, and LLM correlations in Python and Java are weakly negative, with the largest magnitudes around $-0.15$ for DSC Java vulnerabilities and $-0.13$ for OpenAI GPT Java high-severity findings. These directions are compatible with RQ3's observation that some LLM vulnerability findings arise from templated logging, debugging, or configuration patterns, but the magnitudes are too small to support a strong claim. Security profiles are better explained by vulnerability type and language-specific rule coverage than by naturalness alone. 

\begin{mainbox}{}
Structural and statistical properties only partially explain the defect and vulnerability profiles. Raw correlations between complexity and defects are mostly size effects: once whole-sample NLOC is controlled for, conventional structural metrics contribute little explanatory signal, and aggregate security outcomes remain largely uncorrelated with complexity. The few stronger residual associations are specific rather than general: DSC's Function/Class/Object cluster in Python and Java reflects a low-complexity scaffolding pattern, while the C Parameter Count--Interface association reflects the trigger mechanics of easily swappable adjacent parameters. Naturalness is more informative about generation style. Human and OpenAI GPT code show weak positive naturalness--defect associations, whereas DSC and Qwen reverse direction in Python because their most predictable outputs are also boilerplate or scaffolding patterns that trigger recurring defects. Overall, code structure alone does not explain the human-vs-AI quality gap; the most informative signals arise when structural metrics, naturalness, defect taxonomy, and language-specific analyzer behavior are interpreted together.
\end{mainbox}

\section{CQBench: An Issue-Prone Code-Quality Benchmark}
\label{sec:benchmark}
To make the study operational beyond the analyzed models, we derive \emph{CQBench}, a static code-quality challenge benchmark for Python, Java, and C. Its purpose is to test whether a new model produces structurally substantive code with fewer analyzer-detected defects and vulnerability patterns than the OpenAI, DeepSeek, and Qwen baselines provided by this study. CQBench deliberately targets difficult tasks; it is not intended to estimate average model quality over the original corpus.

\paragraph{\textbf{Construction.}}
The construction unit is the source task, identified throughout by its stable dataset key. We begin from the per-task tables used in RQ4 and retain only tasks with complete human and three-model measurements. This leaves 255,879 Python, 193,837 Java, and 177,611 C source tasks before challenge-set selection. 

We apply three gates. First, a model output is \emph{complexity-qualified} when either its mean function NLOC or its mean Halstead Volume is at least 10\% of the corresponding human-reference value. The gate is intentionally permissive: it detects empty or nearly empty implementations without requiring generated code to reproduce human complexity. Second, at least two complexity-qualified models must each have at least three included findings. Third, those same models must share at least one defect type or vulnerability class: an ODC category or normalized CWE. Defect counts are computed only after applying the study's Pylint, PMD, and Clang-Tidy exclusion lists and ODC mappings. Finally, the human reference must be parseable, expose the requested signature, and be structurally nontrivial. 

The resulting benchmark contains 27,346 tasks: 10,354 Python, 10,103 Java, and 6,889 C. Of these, 13,693 are selected through defect-type consensus, 3,176 through CWE consensus, and 10,477 through both. The benchmark is not capped at a predetermined top-$N$: every task satisfying the fixed gates is retained. Each task exposes a canonical prompt containing the language, required signature, and original natural-language specification. Human, OpenAI GPT, DeepSeek, and Qwen outputs and keyed results are distributed as baselines. For C, the unified OpenAI baseline uses GPT-OSS; for Python and Java it uses ChatGPT.

\emph{CQBench} is released with JSONL tasks and baselines, frozen analyzer rules, integrity hashes, native and Docker execution paths, and scripts for validation, evaluation, comparison, and reporting. A user can therefore generate one completion per task, run the same static pipeline, and compare task-paired quality rates with the supplied baselines.

\paragraph{\textbf{Scope and limitations.}}
\emph{CQBench} measures analyzer-detected properties, not functional correctness, semantic equivalence, or exploitability. The complexity gate is a safeguard against degenerate implementations rather than evidence of correctness. Moreover, \emph{CQBench} is failure-derived: selection uses findings from the baseline models against which new systems may be compared. Their observed failure rates are consequently enriched by construction, and a newly evaluated model has not participated in selection. We therefore interpret \emph{CQBench} as an adversarial robustness test over known issue-prone tasks and do not generalize its rates to the original task population. Analyzer coverage, ODC/CWE mappings, and the chosen three-finding and 10\% complexity thresholds remain measurement choices; the frozen rules, manifests, and manual audit make those choices explicit and reproducible.

\begin{table}[t]
\caption{CQBench demonstration: task-paired analyzer outcomes for the human reference and Claude Opus 4.8 on the 600-task evaluation subset (200 tasks per language). Defective and Vulnerable report the share of tasks with at least one defect or Semgrep finding under the frozen pipeline; High Severity the share with at least one Critical or Error rated finding; Clean the strict clean rate, \ie a parseable, signature-preserving, complexity-qualified output with no included findings.}
\label{tab:cqbench_claude}
\centering
\small
\begin{tabular}{llrrrrr}
\toprule
Language & Author & Defective (\%) & Vulnerable (\%) & High Severity (\%) & Clean (\%) & Total Defects \\
\midrule
\multirow{2}{*}{Python} & Human  & 62.0 & 15.5 & 1.5  & 31.5 & 284 \\
                        & Claude & 63.0 & 28.0 & 7.5  & 27.5 & 239 \\
\midrule
\multirow{2}{*}{Java}   & Human  & 57.5 & 12.5 & 2.5  & 39.0 & 268 \\
                        & Claude & 64.5 & 16.0 & 2.5  & 32.0 & 313 \\
\midrule
\multirow{2}{*}{C}      & Human  & 72.0 & 40.0 & 15.0 & 17.0 & 395 \\
                        & Claude & 68.5 & 46.5 & 14.0 & 14.0 & 381 \\
\bottomrule
\end{tabular}
\vspace{-0.5cm}
\end{table}

\paragraph{\textbf{Demonstration: evaluating a post-construction frontier model.}}
To demonstrate that \emph{CQBench} remains challenging for newer and stronger models, we evaluated Claude Opus 4.8, a frontier model released in May 2026. We sampled 600 tasks, 200 per language, deterministically and proportionally across the three consensus strata (294 defect, 76 CWE, 230 mixed, matching the full benchmark), generated one completion per task from the canonical \emph{CQBench} prompt with no auxiliary instructions or generation constraints, and scored all outputs using the same methodology described in the study. The subset is difficult by design: the human references themselves trigger at least one defect on 62.0\% of Python, 57.5\% of Java, and 72.0\% of C tasks. \tableautorefname~\ref{tab:cqbench_claude} reports the paired outcomes.

Claude does not exhibit the same structural compression as other AI authors. Its mean NLOC reaches 98.7\% (Python), 94.9\% (Java), and 78.7\% (C) of the human reference on the same tasks, and its mean Halstead Volume 86.3\%, 91.9\%, and 75.5\%. Claude's defect incidence tracks the human reference in every language, falling below it in C (68.5\% against 72.0\%), and it produces 933 total findings against 947 for Human over the 600 tasks. Its ODC composition mirrors the human profile, with a small Function/Class/Object share and no trace of the skew caused by templated code.

Security is where differences persist. Claude's vulnerability incidence over the 600 tasks is 30.2\% against 22.7\% for Human; the gap is significant in Python (28.0\% against 15.5\%), at the boundary of significance in C (46.5\% against 40.0\%), and not significant in Java (16.0\% against 12.5\%). High-severity incidence is significantly elevated only in Python (7.5\% against 1.5\%, driven by CWE-611 and CWE-78 findings), with Java (2.5\% against 2.5\%) and C (14.0\% against 15.0\%) indistinguishable. The surpluses fall in the classes previously identified as LLM-shifted risk, OS command injection (CWE-78, 26 Python findings against Human 8) and race conditions in C (CWE-362, 18 against 9), while on the memory-safety classes that dominate C findings Claude sits near the human reference.

\emph{CQBench}'s selection is derived from failures of the models available at construction time, so a newly evaluated model has not participated in selection and its failure rates on these tasks are not guaranteed to be elevated. The Claude evaluation shows that the selection generalizes: on 62.8\% of the 600 tasks (65.5\% Python, 52.0\% Java, 71.0\% C), Claude triggers at least one finding in the task's own consensus class, the ODC defect type or normalized CWE class fixed at selection time, and among the 453 tasks where it triggers any finding, 83.2\% include a consensus-class finding (57.3\% of ODC-gate tasks reproduce a consensus defect type; 47.7\% of CWE-gate tasks a consensus CWE). The tasks therefore elicit not merely findings, but the same classes of findings, from a model outside the selection.

\section{Discussion}
\label{sec:discussion}
This study set out to characterize not whether large language models can produce functionally plausible code, but how the code they produce differs from human-written code when both are analyzed at scale under comparable conditions. The four research questions converge on a coherent picture. AI-generated code is structurally simpler and statistically more templated than human-written code (RQ$_1$); it exhibits defect and vulnerability profiles that differ from human-written code not only in frequency, but also in type and severity (RQ$_2$ and RQ$_3$); and these differences are only weakly explained by the conventional structural metrics on which much software-quality tooling relies (RQ$_4$).

\paragraph{\textbf{Structural simplicity is not the same as maintainability.}} Across Python, Java, and C, all LLMs produce functions that are smaller and less complex than human-written ones. This leads to higher Maintainability Index scores for AI-generated code, but this advantage should be interpreted carefully. MI is derived from size, cyclomatic complexity, and Halstead volume; therefore, shorter and less branching code receives better scores mechanically. In our results, this structural compression often coexists with missing checks, unused parameters, scaffolding, hardcoded values, or security-sensitive patterns. Thus, the apparent maintainability advantage of AI code is better understood as a consequence of compactness rather than as evidence of better engineering. The naturalness results reinforce this interpretation. After identifier and literal normalization, AI-generated code remains more predictable than human-written code and clusters closer to other AI-generated code than to the human baseline. This suggests that LLMs introduce not only shorter code, but also a more templated coding style. Such regularity can be useful for detection or benchmarking, but it may also indicate repeated generation patterns that carry repeated defects.

\paragraph{\textbf{Quality differences are model- and language-dependent.}} Human and AI code differ not only in the number of issues they trigger, but also in the types of issues they concentrate. Human-written code tends to contain findings typical of mature and evolved codebases, such as higher control-flow complexity, exception-handling patterns, protected-member access, and memory-safety issues in C. AI-generated code more often concentrates simpler and more repetitive patterns, including unused parameters, unused structures, class-like scaffolding, unchecked return values, hardcoded credentials, and debugging or logging artifacts. These differences are graded across models. OpenAI-family code is generally closest to the human baseline, while DeepSeek-Coder shows the strongest departures in structure, defects, and vulnerabilities. Qwen usually occupies an intermediate position. This means that ``AI-generated code'' should not be treated as a single homogeneous category: model choice affects the kind of quality risks that downstream users inherit. The results are also strongly language-dependent. In Python and Java, LLM-generated code more often triggers vulnerability findings than human code, especially around injection, insecure configuration, information exposure, and hardcoded credentials. In C, however, the trend changes: human-written code contains more high-severity memory-safety findings, while LLM-generated code reduces some buffer-related warnings but shifts risk toward other categories. This is an important practical caveat. AI-generated code is not simply more or less secure than human-written code; its security profile depends on the language and on the dominant vulnerability mechanisms of that ecosystem. 

\paragraph{\textbf{Complexity metrics do not explain the quality gap.}} RQ4 shows that conventional structural metrics are poor standalone explanations of the defect and vulnerability differences observed in RQ2 and RQ3. Raw correlations between complexity and defects are largely size effects: larger samples contain more code and therefore more opportunities for quality issues. Once NLOC is controlled for, most associations between complexity metrics and aggregate defects disappear, and security outcomes remain largely uncorrelated with structure. The few stronger residual associations are narrow and rule-dependent. For example, DeepSeek-Coder exhibits a Python/Java Function/Class/Object pattern consistent with low-complexity scaffolding, while the C association between Parameter Count and Interface defects is partly conditioned by the mechanics of \texttt{bugprone-easily-swappable-parameters}. Naturalness is more informative than structural complexity, but it is also model-dependent: for human and OpenAI-family code, less predictable code is weakly associated with more defects, while for DeepSeek-Coder and Qwen in Python the sign reverses because their most predictable outputs often correspond to boilerplate patterns that trigger recurring defects. The broader methodological lesson is that no single metric family is sufficient. Complexity, naturalness, defect categories, vulnerability types, and analyzer behavior must be interpreted together. 

\paragraph{\textbf{Implications.}} 
For researchers, the results motivate paired, taxonomy-grounded, multi-language studies of generated-code quality. Benchmark pass rates remain useful, but they do not capture how generated code differs from human-written software in maintainability, reliability, and security. Mapping tool-specific findings to ODC and CWE makes results more comparable across languages and analyzers, and avoids reducing the analysis to raw static-analysis rule counts. Future studies should also report results per language and per model, because aggregate ``human versus AI'' claims hide important variation. For practitioners, the results suggest that reviewing AI-generated code requires different priorities from reviewing human-written code. Since generated code is often compact and regular, reviewers should not focus only on excessive complexity. They should instead look for missing validation, unhandled edge cases, unchecked return values, unused parameters, hardcoded values, leaked debug logic, and insecure configuration. Security review should also be language-specific: Python and Java require attention to injection, information exposure, and credentials, while C requires continued attention to memory safety as well as concurrency and command-injection patterns. For tool builders and model developers, the findings point to the need for AI-aware quality checks. Static analyzers and maintainability dashboards calibrated on human code may miss recurring generated-code patterns or overvalue compactness. Evaluation and post-training objectives should reward defensive programming, error handling, secure defaults, and language-specific safety practices, not only functional correctness or brevity. Finally, the results raise a longer-term concern. If AI-generated code is systematically more compact, more templated, and statistically closer to other AI-generated code than to human-written code, widespread adoption may gradually reshape the structure and style of open-source software. Understanding whether such patterns persist, accumulate as technical debt, or re-enter future training corpora is an important direction for longitudinal research.

\section{Threats to Validity}
\label{sec:threats}
\textbf{Construct Validity.}
Our analysis relies exclusively on static analysis to characterize code quality, which restricts the evaluation to syntactic and structural properties and excludes runtime behavior, context-specific errors, and deeper semantic correctness. This choice is dictated by the setting: LLM-generated samples are standalone functions or snippets lacking executable context, build systems, or callers, which makes dynamic approaches such as runtime verification and testing impractical at corpus scale. We mitigate this by using mature, widely adopted analyzers (Pylint, PMD, Clang-Tidy, and Semgrep) that produce consistent, fine-grained, rule-level findings suitable for large-scale comparison. We performed a randomized screening as described in \S~\ref{sec:manual-validation}. Even if tool-level false-positive rates were to propagate into absolute counts, they do not, however, affect the within-tool, within-language comparisons on which our conclusions rest, except where rule precision interacts with author-specific style, a residual threat we flag per finding where relevant.

A related and more fundamental limitation concerns functional correctness. There is no tractable, reliable way to verify at this scale that a generated function actually implements the behavior described by its docstring. Establishing this would require executable oracles, test suites, or reference implementations for nearly 800,000 functions across three languages, none of which exist for arbitrary real-world code drawn from open-source repositories. Our study therefore does not claim to measure whether generated code is correct; it characterizes structural, stylistic, defect, and security \emph{properties} of the code as written. We treat the published functional benchmarks of the four generators (\eg HumanEval, MBPP, MultiPL-E; see \S\ref{sec:setup}) as external evidence that these models produce functional code in general, and use our metrics as proxies for quality dimensions rather than as correctness guarantees. Interpreting our findings as statements about non-functional quality, conditioned on the models' documented functional ability, is the intended reading; interpreting them as evidence about correctness is not.

Because the OpenAI GPT column aggregates two distinct models (ChatGPT/gpt-3.5-turbo for Python and Java, gpt-oss-20B for C), we caution against reading it as a single homogeneous author. Our calibration study (\S\ref{sec:setup-calibration}) shows that the two models are not interchangeable: they differ on complexity and issue measurements. The label denotes shared provenance, not equivalence. All values under this heading are interpreted per language, and the only cross-language claim we draw from it is the preserved \emph{ordering} of authors (the OpenAI entry is consistently the most \textit{human-like} LLM), not the magnitude of any gap.

Mapping heterogeneous static-analysis rules onto a shared taxonomy also involves interpretive judgment, since rule definitions differ across analyzers in both granularity and intent. To reduce subjectivity in the rule-to-ODC mapping, we followed the protocol described in \S\ref{sec:research_study}, with independent classification and consensus resolution of disagreements. Some residual ambiguity remains for rules that cover multiple defect types. The complete mapping is released in the replication package~\cite{replication} for inspection and reuse. For security, we rely on Semgrep's curated rule sets and its native rule-to-CWE metadata, used as-is; coverage is therefore bounded by Semgrep's rule base, and severity reflects Semgrep's own assignment rather than an independent exploitability assessment.

\medskip
\noindent
\textbf{Internal Validity.}
For Python and Java our dataset builds on the publicly available, peer-reviewed \textit{HMCorp} dataset~\cite{xu2025distinguishing}, which pairs human-written and ChatGPT-generated functions; for C, which HMCorp does not cover, we constructed a new corpus from TheVault~\cite{nguyen2023vault} and generated implementations with gpt-oss-20B, DeepSeek-Coder, and Qwen2.5-Coder. Expanding the corpus with additional generators introduces potential variability: differences in model behavior and prompt interpretation could influence the generated code. We mitigate this by \textit{(i)} using the prompting format recommended by each model's authors while keeping the docstring-plus-signature input fixed across models, \textit{(ii)} applying a uniform post-generation cleanup pipeline, and \textit{(iii)} anchoring the human baseline on established, provenance-traceable datasets.

A related threat is training-data contamination. The human functions originate from public GitHub repositories via CodeSearchNet and TheVault, which plausibly overlap with the (undisclosed) training corpora of all four generators; a model may therefore have seen the original implementation of a function it is asked to regenerate. We cannot verify membership, but we note that the direction of the resulting bias is conservative for our conclusions: memorization would pull generated code toward the human reference and shrink the human--AI differences we measure, whereas we observe large systematic gaps in structure, style, defects, and vulnerabilities. Contamination is also partly constitutive of the setting we study, since production use of these assistants likewise involves models trained on public code. Finally, the naturalness analysis is insulated by design: as discussed in \S\ref{sec:naturalness-design}, we score predictability with $n$-gram models trained per author on held-out folds of our own corpus, precisely to avoid the overlap between evaluated code and the pretraining data of any neural scorer. 

A specific generation constraint concerns the OpenAI models. The ChatGPT (gpt-3.5-turbo) implementations for Python and Java were inherited from HMCorp and could not be regenerated or extended to C, as that model was not available to us under the same conditions used to build HMCorp. For C we therefore used gpt-oss-20B, an open-weight model from the same provider. This is why the OpenAI GPT column is model-heterogeneous across languages, and it is the reason we ran the calibration study and scope our cross-language claims to author ordering rather than magnitude.

We generated code with a maximum output length of 512 tokens. While this budget comfortably exceeds the mean human function length in every language (\S\ref{sec:dataset}), it can truncate the longest generations. We report incorrect-sample rates explicitly per author and language and exclude unparseable functions from the metric analyses.

\medskip
\noindent
\textbf{External Validity.}
Our findings are conditioned on the selected programming languages (Python, Java, C), the four generators (the OpenAI GPT models, DeepSeek-Coder, Qwen2.5-Coder), and a docstring-driven generation strategy. Results could differ for other languages, other or larger models, agentic or multi-turn generation, or prompting strategies richer than a single docstring and signature. We chose these configurations for their prevalence in research and practice, and we broadened coverage by considering different and heterogeneous languages, including a systems language with a distinct memory-safety profile; the corpus spans 787,562 functions from roughly 34k real-world GitHub repositories. Even so, generalization beyond these settings is not guaranteed and is an explicit direction for future work.

A further limitation is cross-language comparability. Because each language is analyzed with a different tool (Pylint, PMD, Clang-Tidy) exposing different numbers and kinds of rules, the \emph{absolute} magnitudes of defect and vulnerability findings are not directly comparable across languages: a higher raw count in one language may reflect a more granular rule set rather than lower quality. The ODC and CWE taxonomies make defect and weakness \emph{types} comparable, and our cross-language claims are correspondingly restricted to directional trends and to language-ecosystem effects (\eg the memory-safety concentration in C, the amplified human-vs-AI gap in more permissive ecosystems, and the reversal of the high-severity security ordering in C) rather than to raw cross-language magnitudes.

\section{Conclusion}
\label{sec:conclusion}
This paper presented a large-scale, taxonomy-grounded comparison of human-written and AI-generated code across Python, Java, and C, built on almost 800k function pairs in which each human implementation is matched by AI implementations generated from the same docstring and signature. The paired design, together with the ODC and CWE taxonomies, allows quality differences to be attributed to authorship rather than to task selection, and to be compared across tools and languages.

The four research questions receive consistent answers. On structure and style (RQ$_1$), AI-generated code is compressed and templated: every model produces functions with roughly half the size and branching of human code, and the naturalness analysis shows that this regularity is structural rather than lexical, with AI code statistically closer to other AI code than to the human baseline. On defects (RQ$_2$), the profiles differ in kind rather than only in volume: human code concentrates the findings of mature, evolved codebases, while AI code concentrates repetitive boilerplate patterns such as unused parameters and class scaffolding, graded across models with DeepSeek-Coder the most extreme and the OpenAI models the closest to human. On security (RQ$_3$), the comparison is language-dependent on both volume and severity: LLMs produce more vulnerable samples and more high-severity findings than humans in Python and Java, while in C the ordering of high-severity findings reverses, as human code carries the memory-safety patterns that analyzers rate as most exploitable. On the relationship between these dimensions (RQ$_4$), size dominates the raw structure--defect associations; once it is controlled for, conventional complexity metrics explain little, and the most informative residual signal is naturalness, whose direction separates authors because the most templated LLM outputs are boilerplate that carries boilerplate defects.

The practical reading is that AI-generated code requires quality assurance calibrated to its own failure modes rather than to those of human code: review attention on missing validation, unused parameters, leaked debug logic, and insecure defaults rather than on excessive complexity, and security assessment that is language-specific rather than uniform. To make such evaluation repeatable beyond the models studied here, we release \emph{CQBench}, a benchmark of 27,346 issue-prone tasks with human and model baselines, analyzer configurations, and a reproducible scoring pipeline, together with the full dataset, taxonomy mappings, and analysis scripts.

Future work extends the study along the axes our threat analysis identifies: agentic and multi-turn generation strategies richer than a single docstring-and-signature prompt, larger and more recent models, additional languages, and correctness-conditioned analyses that relate the quality signatures observed here to verified functional behavior. A longitudinal question also remains open: whether the compact, templated style of AI-generated code persists in repositories, accumulates as technical debt, and re-enters the training corpora of future models.

\section*{Data Availability}
The complete replication package, including the paired corpus, analysis scripts, taxonomy mappings, analyzer configurations, and aggregated results, together with the CQBench benchmark is publicly available at \url{https://github.com/dessertlab/CQBench}~\cite{replication}.

\bibliographystyle{ACM-Reference-Format}
\bibliography{bibliography}

\end{document}